\documentclass[sigconf,nonacm]{acmart}

\usepackage{xcolor}
\usepackage{soul}
\usepackage{enumitem}

\usepackage{booktabs}
\usepackage{multirow}
\usepackage[table]{xcolor}

\usepackage{subcaption}
\usepackage{placeins}
\usepackage{dblfloatfix}

\usepackage{tikz}
\usepackage{pgfplots}
\usepgfplotslibrary{groupplots}
\usetikzlibrary{calc}

\pgfplotsset{compat=1.18}

\definecolor{pumilblue}{RGB}{0,114,178}
\definecolor{queueorange}{RGB}{230,159,0}
\definecolor{maxcsggreen}{RGB}{0,158,115}

\usepackage{pgfplots}
\usepgfplotslibrary{groupplots}
\pgfplotsset{compat=1.18}

\pgfplotsset{
    colormap={rankgain}{
        rgb255(0cm)=(165,0,38)
        rgb255(1cm)=(247,247,247)
        rgb255(2cm)=(5,48,97)
    }
}

\pgfplotsset{
    pumil/.style={
        color=pumilblue,
        line width=1.45pt,
        solid,
        mark=*,
        mark size=1.45pt,
        mark options={
            solid,
            fill=pumilblue,
            draw=pumilblue
        }
    },
    queue/.style={
        color=queueorange,
        line width=1.35pt,
        dashed,
        mark=square*,
        mark size=1.40pt,
        mark options={
            solid,
            fill=white,
            draw=queueorange,
            line width=0.8pt
        }
    },
    maxcsg/.style={
        color=maxcsggreen,
        line width=1.35pt,
        densely dotted,
        mark=triangle*,
        mark size=1.60pt,
        mark options={
            solid,
            fill=maxcsggreen,
            draw=maxcsggreen
        }
    }
}

\newcommand{\pumilsymbol}{%
    \protect\tikz[baseline=-0.55ex]{
        \draw[blue, solid, line width=0.7pt]
            (0,0) -- (0.42,0);
        \fill[blue]
            (0.21,0) circle[radius=1.25pt];
    }%
}

\newcommand{\queuesymbol}{%
    \protect\tikz[baseline=-0.55ex]{
        \draw[orange, dashed, line width=0.7pt]
            (0,0) -- (0.42,0);
        \draw[orange, fill=white, line width=0.7pt]
            (0.18,-0.035) rectangle (0.24,0.035);
    }%
}

\newcommand{\maxcsgsymbol}{%
    \protect\tikz[baseline=-0.55ex]{
        \draw[green!60!black, dotted, line width=0.7pt]
            (0,0) -- (0.42,0);
        \fill[green!60!black]
            (0.21,0.045) --
            (0.17,-0.035) --
            (0.25,-0.035) -- cycle;
    }%
}

\AtBeginDocument{%
  }

\setcopyright{none}
\acmConference[Preprint]{Preprint}{October 2026}{}

\begin{document}

\title[Unified CSA Framework for Alert Validation]{Integrating Perception, Comprehension, and Projection for Alert Validation: A Unified Cyber Situational Awareness Framework}

\author{Navodika Karunasingha} 
\correspondingauthor 
\email{n.karunasingha@unsw.edu.au} 
\authornotemark[1]
\affiliation{%
\institution{School of Computer Science and Engineering, UNSW Sydney} 
\country{Australia} 
}

\author{Mohan Baruwal Chhetri} 
\correspondingauthor 
\email{mohan.baruwalchhetri@csiro.au}
\affiliation{%
\institution{CSIRO}
\country{Australia} 
}

\author{Surya Nepal} \correspondingauthor 
\email{surya.nepal@unsw.edu.au} 
\email{surya.nepal@csiro.au} 

\affiliation{
\institution{School of Computer Science and Engineering, UNSW Sydney} 
\country{Australia} 
} 

\affiliation{%
\institution{CSIRO} 
\country{Australia} 
}

\author{Cecile Paris} 
\correspondingauthor 
\email{cecile.paris@csiro.au} 
\affiliation{%
\institution{CSIRO}
\country{Australia}
} 

\author{Salil S. Kanhere} 
\correspondingauthor
\email{salil.kanhere@unsw.edu.au}
\affiliation{%
\institution{School of Computer Science and Engineering, UNSW Sydney}
\country{Australia}
}


\renewcommand{\shortauthors}{Karunasingha et al.}

\begin{abstract}

Security Operation Centres (SOCs) must validate large volumes of alerts with limited analyst resources, yet alerts often lack sufficient context to determine whether suspicious activity is attack-related. While provenance-based intrusion detection research has progressed from perception-level signals to comprehension- and projection-level cyber situational awareness (CSA), it remains unclear whether richer CSA consistently improves alert validation. This paper presents the first systematic evaluation of \mbox{perception-,} 
\mbox{comprehension-,} and projection-level CSA within a unified alert validation framework. Using \textsc{Captain} and \textsc{Kairos}, we generate multi-level CSA and formulate alert validation as a positive--unlabelled multi-instance learning (PU-MIL) task over CSA-derived behavioural hypotheses. Experimental evaluation on the DARPA TC dataset shows that comprehension-level CSA substantially improves validation by incorporating causal attack context. Relative to the Queue and Max. CSG baselines, PU-MIL achieves up to 28.6 and 75 percentage-point gains in Recall@5, and 24.6\% and 86.8\% reductions in mean positive rank on Cadets and Theia, respectively. Projection-level CSA can yield additional benefits, reducing the mean positive rank on Theia from 22.0 to 8.5 at a 40-minute projection horizon, although its effectiveness varies across projection horizons. Overall, richer CSA does not necessarily lead to better alert validation. Rather, its value depends on when and how it is introduced, motivating adaptive CSA progression that selectively enriches alerts according to validation requirements.

\end{abstract}




\maketitle

\section{Introduction} \label{sec:introduction}

Security Operation Centres (SOCs) rely heavily on \textit{intrusion detection systems} (IDSs) to identify early signs of malicious activity in enterprise environments. Beyond raising alerts, IDSs underpin cyber situational awareness (CSA), the analyst's ability to determine whether an observed event is malicious or benign, and to reason about its type, source, and target~\cite{karunasingha2025sok}. Following Endsley’s model, CSA is commonly organised into three levels of increasing analytical depth: \textit{perception}, \textit{comprehension}, and \textit{projection}~\cite{endsley2017toward, barford2009cyber}.

The growing sophistication of multi-stage attacks, coupled with the availability of fine-grained host telemetry, has driven intrusion detection toward system-level provenance analysis, where system execution is modelled as causally connected interactions among system entities such as processes, files, and network \mbox{sockets~\cite{zipperle2022provenance,li2021threat}}. Building on this representation, \textit{provenance-based IDS} (PIDSs) incorporate causal reasoning into alert generation by propagating suspiciousness across these interactions, enabling more context-aware detection and reducing false positives~\cite{hossain2017sleuth,hossain2020combating,li2025incorporating}.

Despite these advances, modern alert generation pipelines continue to produce overwhelming volumes of alerts, many of which are false positives, requiring Tier-I analysts to manually validate each alert before escalation or response. Prior work shows that excessive alert volumes, coupled with limited contextual and causal attribution, significantly increase analyst workload and contribute to alert fatigue in SOCs~\cite{li2021threat,zipperle2022provenance}. Because only a small fraction of generated alerts can realistically be investigated in operational SOCs, alert validation and prioritisation are closely intertwined. Under limited resources, analysts must determine which alerts warrant investigation and should therefore be prioritised.

Validating a suspicious alert requires gathering related evidence and reconstructing its causal context. At the system level, this means correlating causally related system activities into connected provenance structures that capture an attack's progression from its root cause through subsequent system interactions~\cite{li2021threat,zipperle2022provenance}. Such provenance dependency graphs (PDGs) reveal the causal chains of activities associated with an alert, supporting causal awareness of how a security event emerged and evolved~\cite{karunasingha2025sok}.

Recent provenance-analysis research has increasingly focused on automating PDG reconstruction from system activities and integrating them into alert generation~\cite{milajerdi2019holmes,milajerdi2019poirot,hassan2019nodoze,cheng2024kairos}. System activities are causally linked to form PDGs, and alerts are generated based on suspiciousness propagated over these structures rather than isolated events. Consequently, alert generation moves beyond perception-level signals and increasingly relies on comprehension-level CSA, where detection decisions are informed by reconstructed attack evolution and causal dependencies among system activities.
More recently, provenance-guided analysis has shifted toward projection-level reasoning, aiming to anticipate future attacker actions and targets~\cite{liu2023causal,zhu2024nip,karunasingha2025sok}. While these approaches primarily focus on estimating future attack states to support projection-level CSA generation, they remain largely decoupled from alert generation and validation workflows. 

Collectively, the literature reveals an evolutionary progression from perception-level CSA, with limited causal correlation support~\cite{hossain2017sleuth,hossain2020combating,li2025incorporating} to subsequent approaches that explicitly reconstruct PDGs for comprehension-level causal attack reconstruction~\cite{milajerdi2019holmes,milajerdi2019poirot,hassan2019nodoze,cheng2024kairos} and projection-oriented future attack-state \mbox{reasoning~\cite{liu2023causal,zhu2024nip}}. 

Yet, despite this progression, existing work focuses primarily on generating these distinct tiers of CSA rather than systematically evaluating their relative utility. Consequently, it remains unclear whether progressively richer CSA information consistently improves validation effectiveness, what marginal benefit each additional level provides, and whether the additional analytical complexity of generating richer CSA is operationally justified in a SOC. Richer CSA provides access to additional contextual and behavioural evidence, but it also introduces greater computational cost, analytical complexity, and potentially irrelevant information. So, we cannot assume that progressing to higher CSA levels will necessarily improve alert validation performance. 

To the best of our knowledge, no prior study has systematically compared perception-, comprehension-, and projection-level CSA within a unified framework to understand their relative contribution to alert validation effectiveness. This motivates a fundamental question: \textbf{How do perception-, comprehension-, and projection-level CSA influence alert validation effectiveness, and what additional value does each CSA level contribute to the validation process?} To answer this, we present a unified framework to incorporate and evaluate progressively richer CSA in alert validation. Alert validation provides a natural setting to directly assess the value of situational awareness information in distinguishing attack-related alerts from benign ones. 

Specifically, this paper makes the following contributions:
\begin{itemize}[topsep=0pt,leftmargin=*]
    \item We formulate alert validation as a progressive CSA-aware decision-making task and present the first systematic evaluation of CSA levels for alert validation. 
    \item We propose a unified framework for progressive CSA generation and alert validation that integrates all three CSA levels through causal and predictive reasoning.
    \item We instantiate the framework using state-of-the-art approaches, operationalising perception with \mbox{\textsc{CAPTAIN}~\cite{li2025incorporating}}, comprehension and projection using \textsc{Kairos}~\cite{cheng2024kairos} and alert validation using positive--unlabelled multi-instance learning (PU-MIL)~\cite{perini2023learning}.
    \item We empirically demonstrate the value of progressively richer CSA using \textit{Recall@K}, \textit{Mean Positive Rank (MPR)}, and \textit{Projection-Induced-Rank-Shift (PIRS)}. 
\end{itemize}
\vspace{-1mm}
\section{Background}

This section introduces the foundational concepts underpinning our approach and motivates alert validation as a progressive CSA-aware decision-making task. We discuss relevant literature throughout the paper, including the problem motivation (Section~\ref{sec:introduction}), framework components (Section~\ref{sec:conceptual-framework}), and alternative realisations (Appendix~\ref{app:alternative_realisation}). 

\subsection{Alert Analysis and Analyst Workflows}
SOCs rely on automated detection systems to identify potential threats in large-scale enterprise environments. These systems generate alerts based on predefined rules, statistical anomalies, or learned patterns. However, in practice, analysts face high alert volumes and limited context, requiring manual investigation to determine whether an alert represents a true incident~\cite{tariq2025alert}.
A key challenge in this workflow is the disconnect between alert generation and cyber situation awareness (CSA). Alerts are typically presented as discrete events, such as suspicious process executions or network connections, without sufficient context about their causal relationships or broader impacts. Consequently, analysts must reconstruct attack scenarios by correlating evidence from multiple data sources, including system logs, network traces, and audit records: a time-consuming and cognitively demanding task, particularly in high-alert environments.
PIDSs aim to address this limitation by modelling system execution as a graph of causal dependencies, enabling analysts to reason about how events are connected~\cite{cheng2024kairos, jiang2025orthrus}. While these systems provide richer context, it remains unclear how progressively richer CSA information influences automated alert validation under weakly supervised settings.

\begin{figure*}[!ht]
    \centering



    \begin{subfigure}[t]{0.94\textwidth}
        \begin{minipage}[c]{0.04\textwidth}
            \centering
            \textbf{(a)}
        \end{minipage}%
        \begin{minipage}[c]{0.95\textwidth}
            \centering
            \includegraphics[width=\linewidth]
            {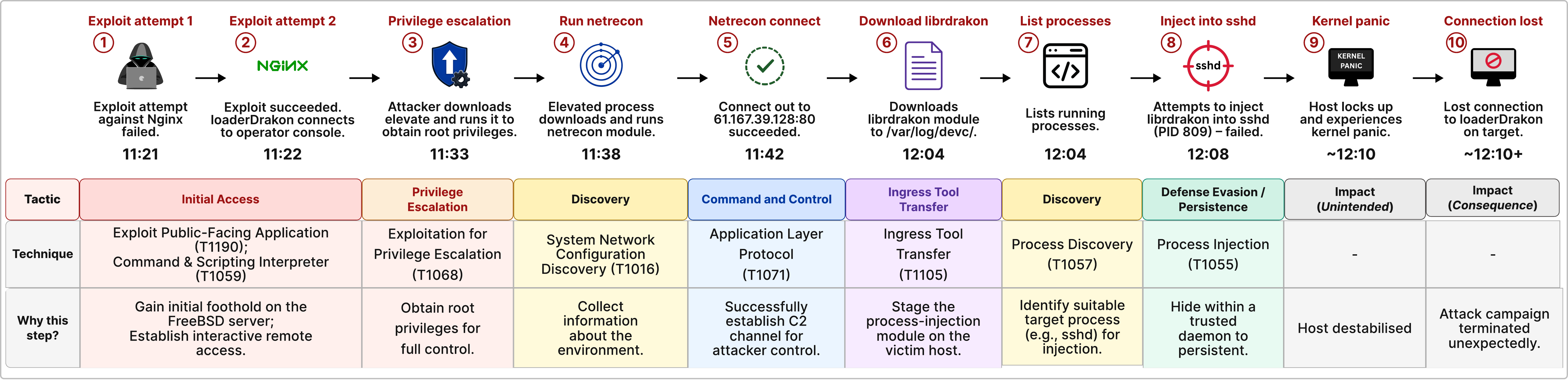}
        \end{minipage}
        \label{fig:motivating_attack_scenario}
    \end{subfigure}



    \begin{subfigure}[t]{0.94\textwidth}
        \begin{minipage}[c]{0.04\textwidth}
            \centering
            \textbf{(b)}
        \end{minipage}%
        \begin{minipage}[c]{0.95\textwidth}
            \centering
            \includegraphics[width=\linewidth]
            {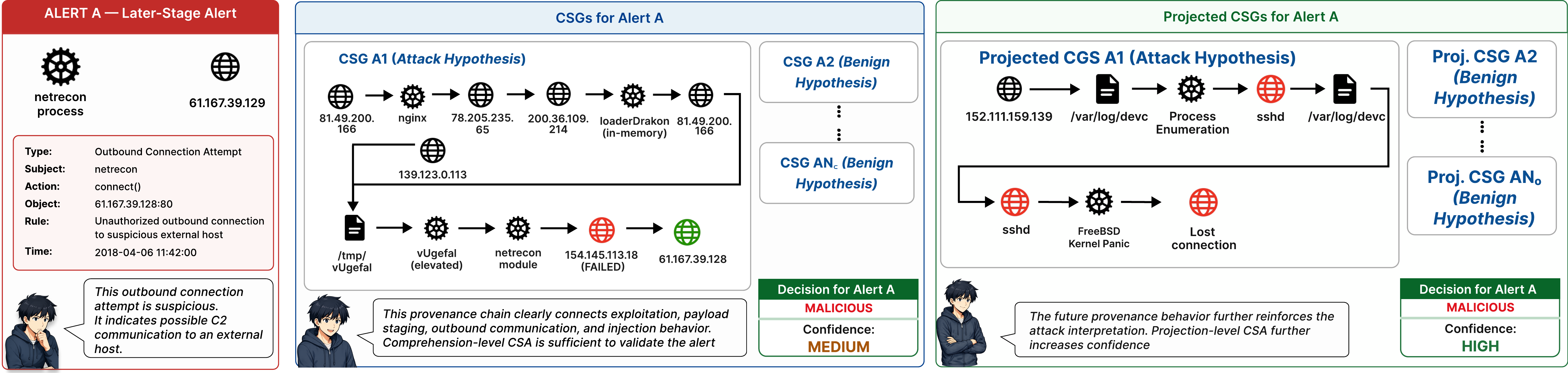}
        \end{minipage}
        \label{fig:motivating_alert_a}
    \end{subfigure}

    \vspace{0.3em}


    \begin{subfigure}[t]{0.94\textwidth}
        \begin{minipage}[c]{0.04\textwidth}
            \centering
            \textbf{(c)}
        \end{minipage}%
        \begin{minipage}[c]{0.95\textwidth}
            \centering
            \includegraphics[width=\linewidth]
            {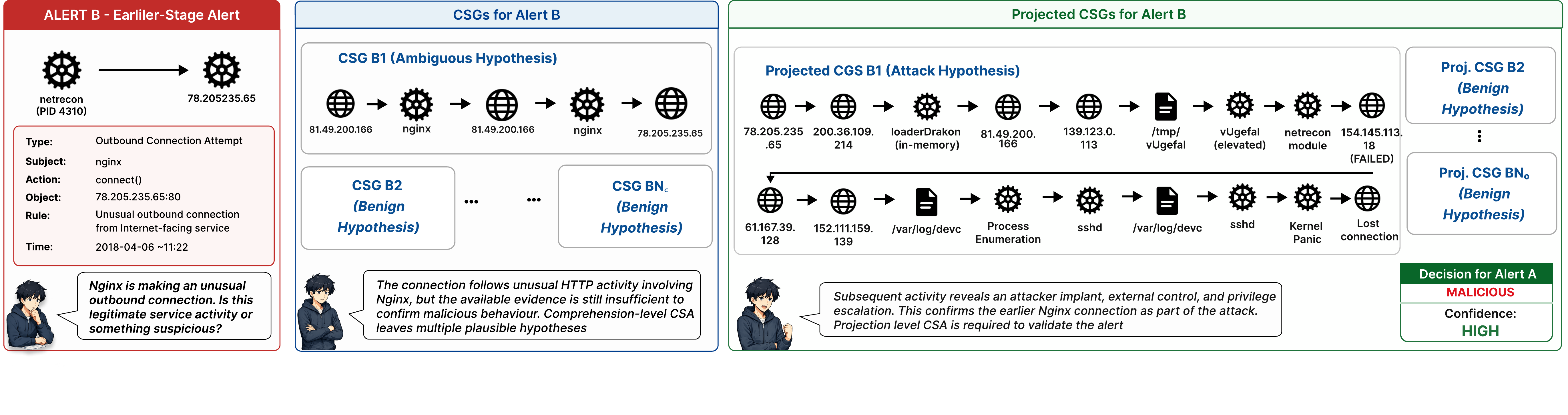}
        \end{minipage}
        \label{fig:motivating_alert_b}
    \end{subfigure}
    \caption{
        Motivating example of progressive CSA-based alert validation using the
        DARPA-TC E3~\footref{fn:darpa-e3} CADETS Day~6 attack campaign. \textbf{(a)} Attack scenario. \textbf{(b)} Validation of later-stage Alert~A. \textbf{(c)} Validation of earlier-stage Alert~B.}
    \label{fig:motivating_example}
\end{figure*}


\subsection{Cyber Situational Awareness}

CSA provides a conceptual framework for how human decision-makers process information in complex environments~\cite{karunasingha2025sok}, commonly structured into three levels: (i) \textit{perception}, which identifies whether the system is in a malicious or benign state by identifying the type, source, and target of a malicious situation; (ii) \textit{comprehension}, which explains the relationships and significance of these elements, such as interpreting causal dependencies between events; and (iii) \textit{projection}, which anticipates future states and potential consequences, including how an attack may evolve. In SOCs, these levels correspond naturally to different forms of information; alerts generated by detection systems correspond to \textit{perception level CSA}~~\cite{hossain2017sleuth,li2025incorporating}, provenance-based summary graphs correspond to \textit{comprehension level CSA}~\cite{cheng2024kairos,jiang2025orthrus}, and predictions or future-oriented context correspond to \textit{projection level CSA}~\cite{liu2023causal,zhu2024nip}. While CSA has been widely studied in domains such as aviation and military operations, its application in cybersecurity, particularly within alert validation workflows, remains underexplored. Existing systems focus primarily on improving detection performance rather than systematically structuring information to support CSA.

\section{Motivating Example} \label{sec:motivating_example}

Figure~\ref{fig:motivating_example} presents a motivating example from the DARPA Transparent Computing (DARPA TC) Cadets dataset\footnote{\url{https://github.com/darpa-i2o/Transparent-Computing}, last accessed May 2026.\label{fn:darpa-e3}}, illustrating a multi-stage attack. The attack begins with exploitation attempts against an Nginx process on a FreeBSD host (\textit{Initial Access}). Following a successful attempt, a connection is established between \texttt{loaderDrakon} and a remote attacker-controlled shell (\textit{Execution, Command and Control}). The attacker downloads and executes malicious modules, including \texttt{netrecon} for outbound communication and attempts to inject \texttt{libdrakon} (\textit{Command and Control, Execution}) into the privileged \texttt{sshd} daemon (\textit{Privilege Escalation, Defense Evasion}). This causes system instability and loss of host connectivity (\textit{Impact}). Alert~A and Alert~B are generated at different stages of the same attack, yet present substantially different behavioural evidence for downstream validation. At the \textit{perception} level, both provide limited CSA (e.g., suspicious entities, triggering interactions, alert metadata), indicating anomalous behaviour without reliably distinguishing malicious from benign activity. 

To enrich this picture, we reconstruct the provenance activity around each alert into a PDG and decompose it into smaller candidate summary graphs (CSGs) 
that represent local behavioural hypotheses~\cite{cheng2024kairos,jiang2025orthrus}. Multiple competing CSGs may be generated for each alert (see Figure~\ref{fig:motivating_example}(b)); only some correspond to the attack progression, while others represent anomalous but benign behaviour. The two alerts differ in how much additional CSA information contributes to alert validation. For Alert~A, \textit{comprehension}-level information is sufficient because the reconstructed CSGs already expose causally connected attack behaviour (e.g., malicious module execution, successful C2 communication, privileged daemon injection); \textit{projection}-level CSA may further increase confidence by revealing the likely consequences (e.g., abnormal \texttt{sshd} behaviour, kernel instability,  communication disruption). Hence, reliable validation is already possible at the \textit{comprehension} level. In contrast, Alert~B remains ambiguous even after \textit{comprehension}-level reconstruction because its CSGs primarily capture suspicious but operationally plausible behaviour (e.g., transient process execution, failed outbound network communication); \textit{projection}-level CSA reveals subsequent attack progression, including successful external communication, malicious payload deployment, daemon injection, and eventual host instability, making the attack path distinguishable from benign alternatives.

This example highlights two challenges: (i) \textit{Alerts from the same attack campaign may require different levels of CSA for reliable validation}. A fixed CSA level can be insufficient for some alerts while introducing unnecessary analysis overhead for others. Here, we focus on the availability of CSA information required for validation, without explicitly modelling human cognitive limitations~\cite{endsley2017toward,jiang2022systematic,tariq2025alert}. (ii) \textit{Provenance reconstruction produces multiple competing CSGs representing alternative behavioural hypotheses}~\cite{cheng2024kairos,jiang2025orthrus}. 
While prior work has identified malicious entities and event labels~\cite{jiang2025orthrus,li2025incorporating}, no existing study, to our knowledge, determines which reconstructed CSGs correspond to the underlying attack path. Alert validation can be formulated as a weakly supervised learning problem in which progressively richer CSA helps identify the most plausible attack-related CSGs among competing hypotheses~\cite{perini2023learning}. This motivates a unified framework to systematically integrate and evaluate CSA across multiple levels.
\section{Framework for Progressive CSA Generation and Alert Validation}
\label{sec:conceptual-framework}
The motivating example in Section \ref{sec:motivating_example} shows that alerts may require different levels of CSA for validation. Some can be confidently validated using preceding context (comprehension-level information); others require projected or observed attack context (projection-level information). Alert validation can therefore be viewed as progressively incorporating CSA into decision-making.

Prior research has explored capabilities such as provenance-based intrusion detection~\cite{hossain2017sleuth,hossain2020combating,li2025incorporating,hassan2019nodoze}, attack reconstruction and provenance summarisation~\cite{cheng2024kairos,jiang2025orthrus}, and attack forecasting~\cite{liu2023causal,zhu2024nip}. Although these capabilities correspond to different levels of CSA, researchers typically evaluate them only against their individual operational objectives. As a result, the systematic generation, integration, and use of progressively richer CSA for alert validation remain largely unexplored.

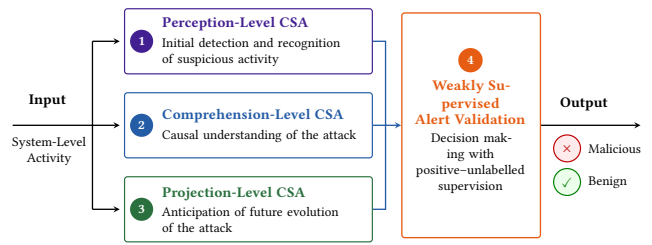
\begin{figure}[!t]
\centering

\resizebox{\columnwidth}{!}{%
\begin{tikzpicture}[
    x=1cm,
    y=1cm,
    font=\normalsize,
    >=stealth
]


\definecolor{perception}{RGB}{67,32,145}
\definecolor{comprehension}{RGB}{36,92,166}
\definecolor{projection}{RGB}{42,112,62}
\definecolor{validation}{RGB}{230,92,18}


\draw[
    -,
    line width=0.9pt
]
(-0.55,0) -- (1.35,0);

\node[
    align=center,
    font=\large\bfseries
] at (0.32,0.58) {
    Input
};

\node[
    align=center,
    text width=1.9cm
] at (0.32,-0.62) {
    System-Level\\
    Activity
};

\coordinate (inputsplit) at (1.35,0);

\draw[
    line width=0.9pt
]
(1.35,-2.0) -- (1.35,2.0);

\draw[
    ->,
    line width=0.9pt
]
(1.35,2.0) -- (2.00,2.0);

\draw[
    ->,
    line width=0.9pt
]
(1.35,0) -- (2.00,0);

\draw[
    ->,
    line width=0.9pt
]
(1.35,-2.0) -- (2.00,-2.0);


\node[
    draw=perception,
    rounded corners=2pt,
    line width=0.8pt,
    minimum width=5.9cm,
    minimum height=1.55cm,
    anchor=west
] (perceptionbox) at (2.10,2.0) {};

\node[
    circle,
    fill=perception,
    text=white,
    font=\bfseries,
    minimum size=0.55cm,
    inner sep=0pt
] at (2.52,2.0) {1};

\node[
    anchor=west,
    align=left,
    text width=4.95cm
] at (2.90,2.0) {
    {\color{perception}\large\bfseries Perception-Level CSA}\\[0.8mm]
    Initial detection and recognition\\
    of suspicious activity
};


\node[
    draw=comprehension,
    rounded corners=2pt,
    line width=0.8pt,
    minimum width=5.9cm,
    minimum height=1.55cm,
    anchor=west
] (comprehensionbox) at (2.10,0) {};

\node[
    circle,
    fill=comprehension,
    text=white,
    font=\bfseries,
    minimum size=0.55cm,
    inner sep=0pt
] at (2.52,0) {2};

\node[
    anchor=west,
    align=left,
    text width=4.95cm
] at (2.90,0) {
    {\color{comprehension}\large\bfseries Comprehension-Level CSA}\\[0.8mm]
    Causal understanding of the attack
};


\node[
    draw=projection,
    rounded corners=2pt,
    line width=0.8pt,
    minimum width=5.9cm,
    minimum height=1.55cm,
    anchor=west
] (projectionbox) at (2.10,-2.0) {};

\node[
    circle,
    fill=projection,
    text=white,
    font=\bfseries,
    minimum size=0.55cm,
    inner sep=0pt
] at (2.52,-2.0) {3};

\node[
    anchor=west,
    align=left,
    text width=4.95cm
] at (2.90,-2.0) {
    {\color{projection}\large\bfseries Projection-Level CSA}\\[0.8mm]
    Anticipation of future evolution\\
    of the attack
};


\coordinate (fusion) at (8.35,0);

\draw[
    comprehension,
    line width=0.9pt
]
(perceptionbox.east) -- ++(0.30,0) |- (fusion);

\draw[
    comprehension,
    line width=0.9pt
]
(comprehensionbox.east) -- (fusion);

\draw[
    comprehension,
    line width=0.9pt
]
(projectionbox.east) -- ++(0.30,0) |- (fusion);


\node[
    draw=validation,
    rounded corners=2pt,
    line width=0.9pt,
    minimum width=3.30cm,
    minimum height=5.35cm,
    align=center,
    anchor=west
] (validationbox) at (8.70,0) {};

\node[
    circle,
    fill=validation,
    text=white,
    font=\bfseries,
    minimum size=0.58cm,
    inner sep=0pt
] at (10.35,1.55) {4};

\node[
    align=center,
    text width=2.85cm
] at (10.35,0.55) {
    {\color{validation}\large\bfseries
    Weakly Supervised\\
    Alert Validation}
};

\node[
    align=center,
    text width=2.85cm
] at (10.35,-0.90) {
    Decision making with\\
    positive--unlabelled\\
    supervision
};


\draw[
    ->,
    comprehension,
    line width=0.9pt
]
(fusion) -- (validationbox.west);


\draw[
    ->,
    line width=0.9pt
]
(validationbox.east) -- (14.40,0);

\node[
    align=center,
    font=\large\bfseries
] at (13.10,0.50) {
    Output
};


\node[
    circle,
    draw=red!75!black,
    fill=red!8,
    line width=0.8pt,
    minimum size=0.64cm,
    inner sep=0pt
] (maliciousicon) at (12.65,-0.55) {
    {\color{red!75!black}\bfseries $\times$}
};

\node[
    anchor=west,
    font=\normalsize
] at (13.05,-0.55) {
    Malicious
};


\node[
    circle,
    draw=green!50!black,
    fill=green!8,
    line width=0.8pt,
    minimum size=0.64cm,
    inner sep=0pt
] (benignicon) at (12.65,-1.35) {
    {\color{green!50!black}\bfseries $\checkmark$}
};

\node[
    anchor=west,
    font=\normalsize
] at (13.05,-1.35) {
    Benign
};

\end{tikzpicture}%
}

\caption{Conceptual framework for progressive CSA generation and alert validation.}
\label{fig:general_overview}

\end{figure}

To address this gap, we conceptualise progressive CSA generation as a modular workflow that incrementally enriches low-level system activity with perception-, comprehension-, and projection-level CSA information prior to alert validation. Each stage contributes distinct contextual evidence, resulting in heterogeneous and variable-sized CSA representations. The lack of fine-grained labels connecting provenance structures to attack progression makes alert validation a weakly supervised problem.
%
Based on this conceptualisation, the proposed framework (see Figure~\ref{fig:general_overview}) comprises four functional stages, discussed below. 

\subsection{Perception-Level CSA Generation} \label{sec:perception_concept}

Perception-level CSA supports \textit{recognition}, which determines whether an observed situation is malicious or benign, and \textit{identification}, which determines the type, source, and target of that situation~\cite{karunasingha2025sok}. In provenance-based security analysis, provenance-based intrusion detection systems (PIDSs) typically generate perception-level CSAs. By analysing causal interactions among entities in provenance graphs, PIDSs identify anomalous or potentially malicious behaviour and generate alerts describing suspicious situations~\cite{cheng2024kairos,hossain2017sleuth,hossain2020combating,li2025incorporating,jiang2025orthrus}. From a CSA perspective, these alerts establish awareness of suspicious activity while identifying the entities and interactions involved.
The increasing adoption of PIDSs has led to diverse approaches for generating perception-level CSA. Although these approaches differ in how they infer suspiciousness, they share the common goal of transforming low-level provenance events into alerts that expose potentially malicious situations. These design choices shape the form and granularity of perception-level CSA, as discussed further in Appendix~\ref{app:perception_realisation}.

\subsection{Comprehension-Level CSA Generation} \label{sec:comprehension_concept}

Comprehension-level CSA supports \textit{causal awareness} by enabling understanding of the causes, mechanisms, and significance of suspicious system behaviour through reconstruction of the causal relationships surrounding an alert~\cite{karunasingha2025sok}. Rather than treating an alert as an isolated event, comprehension-level CSA enriches it with provenance context that captures interactions among system entities and the causal progression of suspicious activities. This enables analysts to reason about how the situation evolved, its underlying causes, and its relationship to the broader attack progression.

In provenance-based security analysis, comprehension-level CSA is typically generated by reconstructing provenance structures surrounding suspicious activity~\cite{king2005backtracking,hassan2019nodoze,milajerdi2019holmes,milajerdi2019poirot,cheng2024kairos}. System execution is represented as causally connected interactions among system entities, from which attack-relevant provenance dependency graphs (PDGs) are  constructed~\cite{king2005backtracking,hassan2019nodoze,milajerdi2019holmes}. These structures capture behavioural relationships associated with an alert and provide causal explanations beyond those available at the perception level.

The increasing adoption of PIDSs has led to diverse approaches for generating comprehension-level CSA. Although they differ in how they reconstruct attack-relevant provenance structures, they share the common objective of transforming low-level provenance activity into causally meaningful behavioural representations. The resulting representations vary in granularity and operational characteristics, and Appendix~\ref{app:comprehension_realisation} discusses different approaches to generating comprehension-level information.

\subsection{Projection-Level CSA Generation} \label{sec:projection_concept}

Projection-level CSA supports reasoning about how suspicious behaviour may evolve beyond the current system state by extending analyst awareness from understanding the present situation to anticipating plausible future developments~\cite{karunasingha2025sok}. Where perception-level CSA identifies suspicious activity and comprehension-level CSA establishes causal awareness of how it emerged, projection-level CSA provides future-oriented context that supports anticipation of the likely progression of attacks, subsequent attacker actions, or system entities yet to be affected.

In provenance-based security analysis, projection-level CSA is commonly generated by reasoning over provenance-derived behavioural representations, such as PDGs or attack graphs, to infer future attack evolution from observed system behaviour~\cite{liu2023causal,zhu2024nip}. These mechanisms extend the reconstructed attack context beyond the currently observed provenance activity, predicting plausible future system states. From a CSA perspective, such predictions constitute projection-level CSA because they provide anticipatory awareness of how an ongoing attack may evolve.

Existing approaches formulate projection-level CSA as a forecasting problem over attack or provenance graphs. Although they differ in how they model and predict future attack behaviour, they share the objective of estimating attack evolution from observed system behaviour. Resulting projections may include anticipated attacker actions, future paths to compromise, or system entities likely to be affected during subsequent attack stages. Appendix~\ref{appendix:projection_generation} discusses alternative realisations of projection-level CSA generation.

\subsection{Progressive CSA-Aware Alert Validation}

Alert validation determines whether a generated alert constitutes a genuine security incident requiring further investigation. Alerts generated by PIDSs rarely provide sufficient CSA to support confident decision-making. Analysts, therefore, retrieve related system activity, reconstruct causal relationships, correlate supporting evidence, and examine subsequent system behaviour. This investigative effort constitutes a substantial component of analyst workload and contributes significantly to alert fatigue in modern SOCs \cite{alahmadi202299,yang2024true}.

Recent research has recognised the value of contextual enrichment for alert validation. Existing approaches augment alerts with historical provenance dependencies, event relationships, and other forms of behavioural context to support triage~\cite{hassan2019nodoze,ede2022deepcase,wang2024combating}. Although these improve downstream decision-making, they generally treat contextual information as a single source of evidence and do not explicitly distinguish between CSA levels or investigate how progressively richer CSA contributes to validation.

Within the proposed framework, alert validation is conceptualised as a progressive process of evidence accumulation. Rather than treating an alert as a static detection outcome, it is progressively enriched with CSA information generated by the upstream stages (Sections~\ref{sec:perception_concept}--\ref{sec:projection_concept}). Validation then evaluates the accumulated evidence, enabling increasingly informed decisions as additional CSA becomes available. 
This formulation reflects operational SOC workflows, where analysts progressively incorporate additional evidence throughout an investigation rather than relying solely on the initial alert. By explicitly organising validation around perception-, \mbox{comprehension-,} and projection-level CSA, the framework provides a structured foundation for automating alert validation while reducing manual investigation effort.

The framework deliberately separates CSA generation from validation, making it independent of any specific algorithm and adaptable to different implementation strategies. Section~\ref{sec:instantiation} presents one concrete instantiation and demonstrates how the generated CSA is operationalised for automated alert validation.

\section{Framework Instantiation} \label{sec:instantiation}

\begin{figure*}[t]
\centering
\includegraphics[
        width=\textwidth,
        keepaspectratio
    ]{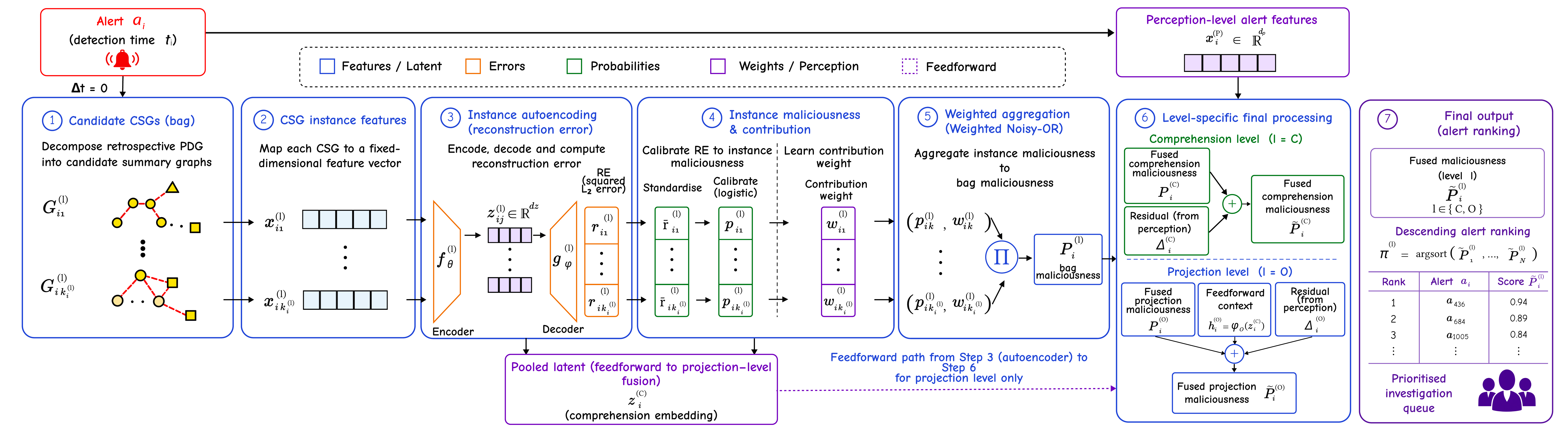}
    \caption{Progressive PU-MIL instantiation for CSA-aware alert validation. CSA information is progressively integrated to estimate alert maliciousness and produce a prioritised investigation queue.}
    \label{fig:system-instantiation}
\end{figure*}

This section provides a concrete instantiation of the framework presented in Section~\ref{sec:conceptual-framework} using selected state-of-the-art provenance-based security analysis techniques (see Appendix \ref{app:alternative_realisation} for alternative instantiations). Rather than introducing new intrusion detection or attack reconstruction mechanisms, this instantiation operationalises the framework through compatible components, enabling a systematic evaluation of how progressively richer CSA influences alert validation. Figure~\ref{fig:system-instantiation} presents the resulting end-to-end system.

\subsection{System Instantiation Overview} \label{sec:system-instantiation-overview}

Figure~\ref{fig:system-instantiation} illustrates the complete system instantiation and the information flow between its constituent components. The workflow begins with chronologically ordered audit records, which are transformed into whole-system provenance graphs that serve as the common input to subsequent stages. \textsc{Captain} analyses the provenance graphs to generate alerts together with their associated perception-level CSA. These alerts then anchor the adapted \textsc{Kairos} pipeline, which reconstructs the provenance context surrounding each alert and generates comprehension-level and projection-level CSGs across different observation horizons.

The outputs of these stages are subsequently transformed into representations suitable for downstream validation. Perception-level CSA contributes alert-centric information; comprehension- and projection-level CSA contribute variable-sized collections of behavioural hypotheses derived from the surrounding provenance context. Together, these heterogeneous representations form the evidence the PU-MIL model consumes.
PU-MIL evaluates this evidence progressively across CSA levels. At the comprehension level, it learns from the behavioural structures available when the alert is generated. The learned comprehension-level representation is then transferred to the projection stage, where it helps interpret subsequent behavioural evolution. The resulting maliciousness estimates rank alerts, producing a prioritised alert queue.


Sections~\ref{sec:perception-level_csa_instantiation}--\ref{sec:projection-level_csa_instantiation} describe the generation and representation of each CSA level, while Figure~\ref{fig:system-instantiation} and   Section~\ref{sec:progressive-validation} present their integration through the progressive PU-MIL validation process.

\subsection{Perception-Level CSA Generation}
\label{sec:perception-level_csa_instantiation}

At the perception level, low-level system activities are transformed into alert-centric CSA. System activities are first modelled as provenance graphs. Nodes represent system entities such as processes, files, sockets, and memory objects; edges capture causal interactions among these entities, including file reads and writes, process execution, process creation, and network communication. Following the PIDS adopted in this work~\cite{li2025incorporating}, confidentiality and integrity tags are propagated through the graph according to predefined information-flow rules. Alerts are generated when the propagated tag values satisfy suspicious flow conditions. 

The resulting alert set \(\mathcal{A}=\{a_1,a_2,\ldots,a_N\}\) forms the basis for perception-level CSA. For each alert, the attributes summarised in Table~\ref{tab:csa_features} are transformed into an alert-level feature representation \(x_i^{(P)}=\phi_P(a_i)\), 
where \(\phi_P(a_i)\) denotes the perception-level feature extraction function. The resulting vector \(x_i^{(P)}\) encodes the information directly available at the time of alert generation, including alert metadata, the participating entities and interactions, and tag-related indicators. As such, it constitutes the perception-level CSA representation of alert \(a_i\) and serves as an input for CSA-aware alert validation.

\begin{table*}[!t]
\centering
\footnotesize
\renewcommand{\arraystretch}{1.12}
\setlength{\tabcolsep}{5pt}

\caption{Feature categories used to instantiate perception-, comprehension-, and projection-level CSA from a security analyst's perspective. They are transformed into fixed-dimensional feature vectors for CSA-aware alert validation.}
\label{tab:csa_features}
\begin{tabular}{p{1.7cm} p{2.8cm} p{12cm}}
\toprule
\textbf{Representation} &
\textbf{Feature Category} &
\textbf{Encoded Information}
\\
\midrule

\multirow{5}{*}{$x_i^{(P)}$}
& Alert type
& Type of suspicious activity inferred from system interactions (e.g., process execution, memory access, file access, and network communication), where the entity's security attributes (e.g., confidentiality and integrity) satisfy the predefined alert-generation policies
\\

& System entities
& Subject and object of the suspicious interaction, where the subject is the acting entity initiating the interaction and the object is the entity affected. Entities include processes, files, sockets, network flows, and other provenance-tracked system objects.
\\

& Entity security attributes
& Integrity and confidentiality tags maintained for the participating system entities during provenance analysis.
\\

& Alert condition satisfaction 
& Indicators of whether the security attributes of the system entities satisfy the alert-generation conditions of the triggered event.
\\


\midrule

\multirow{8}{*}{$x_{ij}^{(C)}$, $x_{ij}^{(O)}$}
& Graph size
& Number of entities and interactions represented in the observed or projected CSG.
\\

& Temporal coverage
& Temporal window spanned by the observed or projected CSG and its relation to the alert detection time.
\\

& Alert-Entity correspondence
& Indicators of whether alert-associated entities appear in the observed or projected CSG.
\\

& Process entities
& Number of process entities in the observed or projected CSG.
\\

& File entities
& Number of file entities in the observed or projected CSG.
\\

& Network entities
& Number of network communication entities in the observed or projected CSG.
\\

& Interaction distribution
& Frequencies of process, file, memory, and network interactions represented in the observed or projected CSG.
\\

& Graph anomaly statistics
& Mean, maximum, and cumulative anomaly scores over the interactions in the observed or projected CSG.



\\

\bottomrule
\end{tabular}
\end{table*}

\subsection{Comprehension-Level CSA Instantiation}

At the comprehension level, each alert is enriched with causally related provenance context by reconstructing attack-relevant provenance structures (e.g., PDGs) surrounding the alert. Specifically, the alert serves as an anchor for retrieving provenance activity from the underlying system execution at the time of the alert generation (\(\Delta t =0)\), enabling retrospective analysis of events contributing to the alert. To improve interpretability and support analyst investigation, we decompose the reconstructed PDG into smaller CSGs using a community detection algorithm (e.g., Louvain), where each CSG represents a locally coherent behavioural hypothesis associated with the alert~\cite{cheng2024kairos}. The resulting comprehension-level CSA for alert \(a_i\) is therefore represented as a collection/bag of CSGs,
\begin{equation}
B_i^{(C)}
=
\left\{
G_{ij}^{(C)}
\right\}_{j=1}^{k_i^{(C)}}
\label{eq:comp_bag}
\end{equation}
where $G_{ij}^{(C)}$ denotes the $j^{th}$ CSG associated with alert $a_i$; $k_i^{(C)}$ is the number of CSGs reconstructed for $a_i$. Since different alerts occur within different provenance contexts, $k_i^{(C)}$ may vary across alerts. 

To incorporate these graph-based representations into downstream progressive CSA-aware alert validation training, we map each CSG to a fixed-dimensional feature vector. Applying $\phi_C(\cdot)$, the comprehension-level feature extraction function, to each \(G_{ij}^{(C)}\) yields $x_{ij}^{(C)}=\phi_C\!\left(G_{ij}^{(C)}\right)$, forming the feature set, 
\begin{equation}
X_i^{(C)}
=
\left\{
x_{ij}^{(C)}
\right\}_{j=1}^{k_i^{(C)}}
\label{eq:comp_feature_set}
\end{equation}
The features in \(X_i^{(C)}\) characterise the structural, temporal, entity-level, interaction-level, and anomaly properties of each CSG, as summarised in Table~\ref{tab:csa_features}.

\subsection{Projection-Level CSA Instantiation}
\label{sec:projection-level_csa_instantiation}

At the projection level, the provenance observation horizon is extended beyond the alert-detection time to incorporate subsequent system behaviour. For an alert $a_i$ detected at $t_i$, the available provenance is expanded by a projection interval $\Delta t$. This produces an extended provenance trace that spans from the initial system interaction at $t_0$ to $t_i+\Delta t$. The additional activities observed within  $(t_i,t_i+\Delta t]$ are incorporated into the same reconstruction and summarisation model used at the comprehension-level. As a result, the CSGs associated with the alert may change, yielding the projection-level CSG collection:
\begin{equation}
B_i^{(O)}
=
\left\{
G_{ij}^{(O)}
\right\}_{j=1}^{k_i^{(O)}}
\label{eq:proj_bag}
\end{equation}
where $G_{ij}^{(O)}$ denotes the $j^{th}$ CSG associated with alert $a_i$ after extending the provenance observation horizon, and $k_i^{(O)}$ is the number of resulting CSGs. Because newly observed activities may alter the reconstructed provenance context, both the composition and the number of CSGs can differ from those obtained at the comprehension level.

As in the comprehension stage, we map each projection-level CSG to a fixed-dimensional feature vector. Applying the graph feature extraction function $\phi_O(\cdot)$ to each $G_{ij}^{(O)}$ yields $x_{ij}^{(O)}=\phi_O\big(G_{ij}^{(O)}\big)$,  giving the projection-level feature set for alert \(a_i\), which is defined as
\begin{equation}
X_i^{(O)}
=
\left\{
x_{ij}^{(O)}
\right\}_{j=1}^{k_i^{(O)}}
\label{eq:proj_feature_set}
\end{equation}
Accordingly, $X_i^{(O)}$ characterises the same feature categories defined for comprehension-level CSGs in Table~\ref{tab:csa_features}. The distinction between comprehension- and projection-level representations therefore arises solely from the provenance horizon over which their CSGs are reconstructed. While comprehension-level CSA captures the behavioural hypotheses available at the time of alert generation, projection-level CSA incorporates the subsequent evolution of those behaviours, providing information with which to anticipate how malicious activity may unfold after an alert is raised.

\subsection{Progressive CSA-Aware Alert Validation}
\label{sec:progressive-validation}

\paragraph{Problem Formulation.} Progressive CSA-aware alert validation determines whether an alert corresponds to genuine malicious activity by leveraging evidence accumulated at different CSA levels: perception, comprehension, and projection. Unlike conventional alert classification, the learning problem exhibits two distinctive characteristics:  the evidence per alert is a variable-sized collection of behavioural hypotheses rather than a single fixed observation, and reliable negative labels are unavailable.

The first characteristic follows the structure of progressive CSA generation. Perception-level CSA provides an alert-level representation \(x_i^{(P)}\), whereas comprehension and projection stages associate the same alert with collections of CSGs, denoted by \(B_i^{(C)}\) and \(B_i^{(O)}\), respectively. Consequently, alert validation at successive CSA levels operates over increasingly richer evidence spaces: \(x_i^{(P)}\) at the perception level, \(\left[{x_i^{(P)}, B_i^{(C)}}\right]\) at the comprehension level, and \(\left[{x_i^{(P)}, B_i^{(C)}, B_i^{(O)}}\right]\) at the projection level. Since \(B_i^{(C)}\) and \(B_i^{(O)}\) may contain multiple CSGs whose cardinalities vary across alerts, an alert cannot be treated as a single independently labelled instance. Moreover, ground truth is available only at the alert level, not for individual CSGs. This motivates a \textit{multi-instance learning} formulation, in which each alert is represented as a bag of CSA-derived evidence, and labels are observed at the bag level~\cite{ilse2018attention,dietterich1997solving,carbonneau2018multiple}.

The second characteristic follows from the nature of the available ground truth. As described in section \ref{sec:dataset}, we use the DARPA-TC dataset together with the interaction-level annotations released by \textsc{Captain}, which provide positive attack-event attribution derived from documented attack kill chains but do not exhaustively verify benign interactions~\cite{li2025incorporating}. A confirmed attack alert thus provides a positive label, whereas an alert without ground-truth attribution cannot be treated as confirmed benign. Thus, the observed label \(y_i \in \{0, 1\}\) indicates whether alert \(a_i\) is positively identified as malicious, not its true class. Progressive CSA-aware alert validation is consequently formulated as a \textit{positive-unlabeled multi-instance learning} (PU-MIL) problem~\cite{perini2023learning,elkan2008learning,kiryo2017positive},aiming to learn an alert-level maliciousness score and latent alert representations from partially labelled, variable-cardinality CSA evidence.

\paragraph{Instance-Level Maliciousness Modelling.} \label{sec:instance-level-maliciousness}
Under the PU-MIL formulation, supervision is available only at the alert level, preventing direct estimation of maliciousness for individual CSA instances. PU-MIL therefore parameterises instance-level maliciousness through reconstruction-based anomaly modelling, leveraging the intuition that behavioural patterns frequently observed in the training data should be reconstructed accurately, whereas rare or weakly represented patterns incur larger reconstruction discrepancies~\cite{perini2023learning}.

Given a CSA-derived feature representation \(x_{ij}^{(l)} \in X_i^{(l)}\) where \(l \in \{C,O\}\), an encoder \(f_{\theta^{(l)}}(\cdot)\) produces a latent representation \(z_{ij}^{(l)}=f_{\theta^{(l)}}\left(x_{ij}^{(l)}\right)\), subsequently reconstructed by a decoder \(g_{\phi(l)}(\cdot)\), where \(\hat{x}_{ij}^{(l)}=g_{\phi(l)}\left(z_{ij}^{(l)}\right)\). Behavioural deviation is quantified using the squared reconstruction error (RE), 
\begin{equation}
    r_{ij}^{(l)}=\left\|x_{ij}^{(l)}-\hat{x}_{ij}^{(l)}\right\|_2^2
    \label{eq:reconstruction_error}
\end{equation}
The REs from Eq.~\ref{eq:reconstruction_error} quantify relative behavioural deviation but admit no direct probabilistic interpretation. To obtain a scale-invariant anomaly measure suitable for probability calibration~\cite{perini2023learning}, the REs are standardised,
\begin{equation}
    \tilde r_{ij}^{(l)}=\frac{r_{ij}^{(l)}-\mu_r}{\sigma_r+\epsilon}
    \label{eq:standardised_re}
\end{equation}
where \(\mu_r\) and \(\sigma_r\) denote the mean and standard deviation of REs observed during training; \(\epsilon\) is a small constant for numerical stability.
The standardised anomaly scores \(\tilde r_{ij}^{(l)}\) are  transformed into instance-level maliciousness probabilities through a learnable logistic calibration function
\begin{equation}
    p_{ij}^{(l)}=\sigma(\alpha \tilde r_{ij}^{(l)}+\beta)
    \label{eq:instant-probability}
\end{equation}
where \(\sigma(\cdot)\) is the sigmoid function and \(\alpha,\beta\) are learnable calibration parameters~\cite{perini2023learning}. This maps reconstruction-based anomaly evidence onto the unit interval while preserving the relative ordering of anomalous instances,enabling probabilistic reasoning within the subsequent PU-MIL modelling.

The resulting probabilities \(\{{p_{ij}}^{(l)}\}_{j=1}^{k_i}\) obtained from Eq.~\ref{eq:instant-probability} constitute the instance-level maliciousness scores for the CSGs associated with alert \(a_i\), providing a differentiable parameterisation of instance-level maliciousness that can subsequently be aggregated into alert-level estimates under bag-level supervision.

\paragraph{Alert-Level Probability Estimation.}
Progressive CSA generation may associate multiple CSGs with the same alert, so we aggregate instance-level maliciousness estimates at the alert level. Under PU-MIL, an alert is malicious whenever at least one associated CSG exhibits malicious behaviour~\cite{ilse2018attention}, so we estimate alert-level maliciousness using a weighted Noisy-OR formulation~\cite{perini2023learning}. The probability that alert \(a_i\) is malicious at CSA level \(l\in\{C,O\}\) is:
\begin{equation}
    P_i^{(l)}
    =
    1-
    \prod_{j=1}^{k_i^{(l)}}
    \left(
    1-p_{ij}^{(l)}
    \right)^{w_{ij}^{(l)}}
    \label{eq:alert-probability}
    \end{equation}
where \(p_{ij}^{(l)}\) is the maliciousness probability of the \(j^{th}\) CSG representation in \(X_i^{(l)}\), and \(w_{ij}^{(l)}\) is its relative contribution to alert-level maliciousness estimation. The weighted product is the probability that all CSGs associated with alert \(a_i\) are benign, while its complement is the probability that at least one CSG exhibits malicious behaviour~\cite{perini2023learning}. Consequently, \(P_i^{(l)}\) provides an alert-level maliciousness estimate consistent with the underlying PU-MIL assumption.

The comprehension-level instance latents are also aggregated into a fixed-dimensional bag representation using the same contribution weights,
\begin{equation}
z_i^{(C)}
=
\sum_{j=1}^{k_i^{(C)}}
w_{ij}^{(C)} z_{ij}^{(C)}
\label{eq:comprehension-bag-representation}
\end{equation}
This \textit{contribution-weighted pooling} preserves each CSG's relative importance when constructing the alert-level representation, yielding $z_i^{(C)}$ as the learned comprehension-level bag representation.

\paragraph{Progressive CSA Fusion.}
Alert validation should preserve the awareness accumulated across preceding CSA levels. PU-MIL estimates an alert-level maliciousness probability \(P_i^{(l)}\) (obtained from Eq.~\ref{eq:alert-probability}) from the CSG representations associated with alert \(a_i\). Although these capture increasingly rich behavioural evidence, they may not fully retain the alert-level information from lower CSA levels. Consequently, perception-level alert features \(x_i^{(P)}\) are incorporated as a residual refinement of the CSG-derived maliciousness estimate.

At the comprehension level, perception-level alert information refines this estimate through a bounded residual correction,
\[
\Delta_i^{(C)}=\lambda_C\tanh\left(h_C\left(x_i^{(P)}\right)\right)
\]
where \(h_c(.)\) is a learnable perception-level fusion function and \(\lambda_C\) controls the maximum correction magnitude. Applying this correction in logit space yields the fused comprehension-level estimate,
\begin{equation}
    \widetilde P_i^{(C)}=\sigma\left(\operatorname{logit}\left(P_i^{(C)}\right)+\Delta_i^{(C)}\right)
    \label{eq:comprehension-probability}
\end{equation}
Projection-level validation must additionally remain consistent with the behavioural understanding established during comprehension-level analysis. This is achieved by conditioning fusion on both  \(x_i^{(P)}\) and \(z_i^{(C)}\), giving the projection-level refinement:,
\[
\Delta_i^{(O)}=\lambda_O\tanh\Big(
h_O\Big(
\Big[x_i^{(P)},z_i^{(C)}\Big]
\Big)
\Big)
\]
Since projection-level CSA typically represents anticipated rather than directly observed behaviour, its reliability may vary across alerts. We account for this through a learned gating function,
\[
g_i^{(O)}=g_O\Big(
\big[x_i^{(P)},z_i^{(C)}\big]
\Big),
\qquad
0 \le g_i^{(O)} \le 1
\]
The resulting projection-level maliciousness estimate is,
\begin{equation}
    \widetilde P_i^{(O)}=\sigma\left(\operatorname{logit}\big(P_i^{(O)}\big)+
    g_i^{(O)}
    \Delta_i^{(O)}
    \right)
    \label{eq:projection-probability}
\end{equation}
Conditioning projection-level fusion on \(z_i^{(C)}\) enables incorporation of behavioural understanding from comprehension-level analysis for projection-level validation. The gating function regulates the influence of projected behavioural evolution according to its estimated reliability.

\paragraph{Positive--Unlabeled Optimisation.}
While reconstruction-based anomaly modelling identifies behavioural deviations, RE alone cannot distinguish malicious activity from rare but benign behaviour. Consequently, optimisation must leverage attack-confirmed alerts to associate anomalous behavioural patterns with malicious activity, without assuming that alerts lacking attack attribution are benign. We distinguish the attack-confirmed alerts \(\mathcal{A}^{+}=\{a_i \mid y_i=1\}\) from the remaining unlabelled alerts \(\mathcal{A}^{U}=\{a_i \mid y_i=0\}\).

Optimisation combines unlabelled reconstruction with supervision from attack-confirmed alerts and dynamically selected reliable negatives~\cite{perini2023learning}. The unlabelled component is defined as,
\begin{equation}
\mathcal{L}_{U}^{(l)}= \frac{1}{|\mathcal{A}^{U}|}
\sum_{a_i \in \mathcal{A}^{U}}
\left(
\frac{1}{k_i^{(l)}}
\sum_{j=1}^{k_i^{(l)}}
r_{ij}^{(l)}
\right)
\label{eq:unlabeled-loss}
\end{equation}
where $r_{ij}^{(l)}$ is the RE of the $j^{th}$ CSG instance for alert $a_i$. Thus, all unlabelled alerts contribute to learning the behavioural distribution through reconstruction, without a negative class label.

To provide negative supervision without treating all unlabelled alerts as benign, a reliable-negative set $\mathcal{R}^{(l)} \subseteq \mathcal{A}^{U}$ is selected at each training epoch. Specifically, the $|\mathcal{A}^{+}|$ unlabelled alerts with the lowest current maliciousness estimates $\widetilde P_i^{(l)}$ are selected, such that \(|\mathcal{R}^{(l)}|=|\mathcal{A}^{+}|\). As the maliciousness estimates evolve during optimisation, the
reliable-negative set is re-selected at each epoch. The corresponding
bag-level supervised objective is,
\begin{equation}
\begin{aligned}
\mathcal{L}_{P}^{(l)}
=&
-
\sum_{a_i \in \mathcal{A}^{+}}
\log
\Big(
\widetilde P_i^{(l)}
\Big)
\\
&
-
\sum_{a_i \in \mathcal{R}^{(l)}}
\log
\Big(
1-\widetilde P_i^{(l)}
\Big)
\\
&
+
\lambda
\big(
\alpha^2+\beta^2
\big)
\end{aligned}
\label{eq:pu-supervised-loss}
\end{equation}
The positive component of Eq.~\ref{eq:pu-supervised-loss} promotes high maliciousness probabilities for attack-confirmed alerts, linking reconstruction-based anomaly evidence with known attack behaviour. The reliable-negative component constrains only the currently lowest-scoring unlabelled alerts towards low maliciousness probabilities, avoiding the assumption that all unlabelled alerts are benign. Regularisation of the calibration parameters \(\alpha\) and \(\beta\) further reduces overfitting during optimisation. The overall objective at CSA level \(l\) is therefore, 
\begin{equation}
    \mathcal{L}^{(l)}=\mathcal{L}_{U}^{(l)}+\mathcal{L}_{P}^{(l)}
    \label{eq:loss-function}
\end{equation}
Although supervision is available only at the alert level, \(\widetilde P_i^{(l)}\) (obtained from Eq.~\ref{eq:comprehension-probability} or Eq.~\ref{eq:projection-probability}) is derived from the underlying instance-level probabilities through the aggregation and fusion stages. Minimising \(\mathcal{L}^{(l)}\) therefore propagates bag-level supervision through these stages, encouraging higher maliciousness probabilities for behavioural patterns associated with attack-confirmed alerts~\cite{perini2023learning}. Consequently, the model learns to distinguish attack-associated anomalous behaviour from the behavioural variation in unlabelled alerts, rather than relying solely on RE magnitude.

\paragraph{Alert Ranking.}
Following optimisation, each alert \(a_i\) is associated with a fused maliciousness estimate \(\widetilde P_i^{(l)}\) (obtained in Eq.~\ref{eq:comprehension-probability} or Eq.~\ref{eq:projection-probability}), which reflects the likelihood that the alert corresponds to malicious activity given the available CSA information. Because analyst resources are limited, alert validation ultimately becomes a prioritisation task: investigate alerts most likely to be malicious first. Alerts are ranked by their fused maliciousness estimates:
\begin{equation}
    \pi^{(l)}
    =\operatorname{argsort}
    \left(
    \widetilde P_1^{(l)},
    \widetilde P_2^{(l)},
    \ldots,
    \widetilde P_N^{(l)}
    \right)
\label{eq:alert_ranking}
\end{equation}
where \(\pi^{(l)}\) denotes the ranking of alerts in descending order of maliciousness probability at CSA level $l$. Alerts with larger \(\widetilde P_i^{(l)}\) receive higher investigation priority. 

\section{Evaluation and Results}

We evaluate the progressive CSA information generation framework for alert validation, examining whether higher levels of CSA information, from perception to comprehension to projection, improve alert-validation effectiveness and how projection-level CSA effectiveness varies with the projection horizon \(\Delta t\).
%
%
The evaluation addresses the following research questions:
\begin{enumerate}[leftmargin=*]
    \item[] \textbf{RQ1:} How does progressively richer CSA information influence alert validation effectiveness?
    \item[] \textbf{RQ2:} How does the projection horizon \(\Delta t \) influence the usefulness of projection-level CSA information for alert validation?
\end{enumerate}

\subsection{Experimental Setup}
\paragraph{Datasets.}
\label{sec:dataset}
Dataset selection is constrained by the systems used to instantiate the framework: \textsc{Captain} generates perception-level CSA, while \textsc{Kairos} provides the provenance analysis used to construct comprehension- and projection-level CSA. Consequently, evaluation requires datasets supported by both systems.
We use the DARPA-TC Engagement~3 (E3)\footnote{DARPA-TC Engagement Data Release, \url{https://github.com/darpa-i2o/Transparent-Computing/blob/master/README-E3.md}, last accessed May 2026.\label{fn:darpa-e3}} and Engagement~5 (E5)\footnote{DARPA-TC Engagement Data Release, \url{https://github.com/darpa-i2o/transparent-computing}, last accessed May 2026.\label{fn:darpa-e5}} datasets, widely used benchmarks for provenance-based security analysis. 
Table~\ref{tab:dataset_selection} summarises the overlap in \textsc{Captain} and \textsc{Kairos} support across engagements and provenance sources. Although \textsc{Cadets-E3}, \textsc{Cadets-E5}, and \textsc{Theia-E3} are compatible with the end-to-end framework, the pretrained \textsc{Captain} model for \textsc{Cadets-E5} is not publicly available. We therefore trained a \textsc{Captain} model
on high-memory computing resources.\footnote{The trained
\textsc{Captain-E5} model will be made publicly available upon
submission to a peer-reviewed conference.} However, the trained model produced only two attack-confirmed alerts, which were too few positive instances to support meaningful PU-MIL training and evaluation. Consequently, we excluded Cadets-E5 from subsequent experiments.

All experiments are therefore conducted on \textsc{Cadets-E3} and \textsc{Theia-E3}.
These datasets provide complementary provenance settings with substantial variation in attack campaigns, benign workloads, scale, entity composition, and event structure, enabling evaluation across heterogeneous settings while remaining compatible with the CSA pipeline.
\begin{table}[!t]
\centering
\caption{DARPA-TC dataset coverage and the datasets selected for end-to-end evaluation.}
\label{tab:dataset_selection}
\small
\setlength{\tabcolsep}{2.5pt}
\begin{tabular}{lccccccc}
\toprule
\multirow{2}{*}{\textbf{Data Source}} &
\multicolumn{2}{c}{\textbf{\textsc{Captain}}} &
\multicolumn{2}{c}{\textbf{\textsc{Kairos}}} &
\multirow{2}{*}{\textbf{Selected}} &
\multirow{2}{*}{\textbf{Train}} &
\multirow{2}{*}{\textbf{Evaluation}} \\
\cmidrule(lr){2-3}
\cmidrule(lr){4-5}
&
\textbf{E3} &
\textbf{E5} &
\textbf{E3} &
\textbf{E5} &
&
&
\\
\midrule
\textsc{Cadets}
& \checkmark & \checkmark
& \checkmark & \checkmark
& E3
& 7--12
& 6 \\

\textsc{ClearScope}
& --
& --
& \checkmark & \checkmark
& --
& --
& -- \\

\textsc{FiveDirections}
& --
& --
& --
& --
& --
& --
& -- \\

\textsc{Theia}
& \checkmark & --
& \checkmark & \checkmark
& E3
& 10--11
& 12 \\

\textsc{Trace}
& \checkmark & \checkmark
& --
& --
& --
& --
& -- \\

\bottomrule
\end{tabular}
\end{table}



\paragraph{Training and Evaluation Protocol.}
We adopt a cross-day protocol, training and evaluating on disjoint provenance structures. Compared with random alert-wise splitting, this tests generalisation to unseen attack executions while preventing structure leakage.

Training and evaluation recordings are summarised in Table~\ref{tab:dataset_selection}. For each dataset, we train a PU-MIL model using CSA information from the training days and evaluate it on the held-out day.
%
%
%
Although DARPA-TC attack campaigns span multiple days, each day is an independent execution of the attack scenario rather than part of a continuous multi-day attack trace. The training days thus provide multiple independent executions for learning behavioural patterns; the held-out day assesses generalisation to an unseen execution.

Comprehension- and projection-level evaluations differ only in the temporal scope of provenance available for CSA generation. Comprehension-level evaluation uses only provenance preceding the alert detection time ($\Delta t = 0$); projection-level evaluation extends the observation window beyond detection by $\Delta t > 0$. Varying $\Delta t$ lets us quantify how increasing future system behaviour affects alert validation, with all other factors held constant.


We modified the protocol for \textsc{Theia E3}: a preliminary analysis showed \textsc{Captain} generating many false-positive alerts outside the documented attack period, substantially increasing provenance reconstruction cost without adding attack-relevant evidence. Therefore, we restricted comprehension- and projection-level evaluation to alerts within the documented attack window. This preprocessing improves computational efficiency only; it does not alter the ground truth or evaluated attack alerts, nor does it introduce information unavailable during alert validation.



Since the original DARPA-TC datasets lack interaction-level ground-truth annotations, we use the interaction-level annotations released with the \textsc{Captain} implementation only for PU-MIL optimisation and final evaluation, not during CSA generation (perception-level alert generation or provenance reconstruction). Consistent with positive--unlabelled learning, we treat only attack-confirmed alerts as positive during optimisation; all others are unlabelled.

\paragraph{Baselines.}
\label{sec:baselines}
As described in Section~\ref{sec:perception-level_csa_instantiation}, alert generation produces an unordered set with no mechanism for estimating the relative likelihood that individual alerts are malicious~\cite{li2025incorporating}. Perception-level CSA alone yields no ranking or validation score comparable to the proposed framework, and therefore cannot serve as an alert validation baseline.
%
%
By contrast, the provenance-based attack reconstruction used for comprehension- and projection-level CSA produces anomaly scores when detecting anomalous activity~\cite{cheng2024kairos}. These are the primary quantitative suspiciousness indicators available before CSA-aware learning, making them appropriate baselines.


These anomaly scores originate from attack-centric provenance reconstruction rather than alert-centric validation. Since provenance reconstruction is anchored on the alert timestamp for both comprehension- and projection-level CSAs, multiple alerts tied to the same attack may share the same anomalous queue and reconstructed CSGs. The scores therefore reflect the suspiciousness of reconstructed attack behaviour rather than the relative maliciousness of individual alerts. PU-MIL addresses this by learning alert-specific validation scores from the combined CSA representations.
\paragraph{Evaluation Metrics.}

We treat alert validation as a ranking problem, reflecting the operational setting where analysts investigate a limited number of alerts in descending order of estimated maliciousness. The primary metric is \textbf{Recall@$K$ (R@$K$)}, the proportion of attack-confirmed alerts within the top-$K$ positions:
\begin{equation}
\mathrm{R@K}
=
\frac{
\left|
\mathcal{A}^{+}
\cap
\mathrm{TopK}\!\left(\pi\right)
\right|
}{
\left|
\mathcal{A}^{+}
\right|
}
\end{equation}

\noindent
where $\mathcal{A}^{+}$ is the set of attack-confirmed alerts and $\mathrm{TopK}(\pi)$ the first $K$ alerts in ranking $\pi$. Higher R@$K$ means more attack-confirmed alerts fall within a fixed investigation budget.
%
%
We report R@$K$ at multiple investigation depths to capture both early-ranking effectiveness and broader alert coverage: smaller $K$ reflects performance under tightly constrained analyst capacity, while larger $K$ shows how quickly the ranking recovers all attack-confirmed alerts.

R@$K$ evaluates ranking only at discrete cut-offs. \textbf{Mean Positive Rank} (MPR) is a complementary measure of overall placement:

\begin{equation}
\mathrm{MPR}
=
\frac{1}
{\left|\mathcal{A}^{+}\right|}
\sum_{a_i \in \mathcal{A}^{+}}
\mathrm{rank}_{\pi}(a_i)
\end{equation}

\noindent
where $\mathrm{rank}_{\pi}(a_i)$ is the position of attack-confirmed alert $a_i$ in the ranked list. Lower MPR means attack-confirmed alerts surface earlier, on average. We use R@$K$ as the primary operational metric for malicious-alert coverage under different budgets, and MPR as a complementary measure of overall ranking quality.


To understand how projection-level CSA affects individual attack-confirmed alerts, we also analyse alert-level rank transitions. For alert $a_i$, the \textbf{Projection-Induced Rank Shift} (PIRS) at projection horizon $\Delta t$ is defined as
\begin{equation}
G_i(\Delta t)
=
r_i(0)-r_i(\Delta t)
\end{equation}
where $r_i(0)$ is the rank under comprehension-level CSA ($\Delta t=0$) and $r_i(\Delta t)$ the rank after incorporating projection-level CSA. A positive value moves the alert closer to the front of the investigation queue; a negative value moves it in the opposite direction. Unlike R@$K$ and MPR, which measure aggregate prioritisation, rank gain shows how individual alerts respond to different $\Delta$.



\subsection{Impact of CSA Richness on Alert Validation}
\label{sec:rq1}

RQ1 examines how progressively richer CSA influences alert validation effectiveness. 
Since perception-level CSA establishes awareness of suspicious situations but provides no alert-specific prioritisation score, our quantitative analysis focuses on the additional evidence introduced at the comprehension and projection levels. 
We examine these contributions from two complementary perspectives: the anomaly-score-based rankings characterise how attack-relevant evidence manifests in the generated provenance context; PU-MIL assesses how effectively the same CSA information can be exploited 
to improve alert validation. This separates the contribution of progressively richer CSA from the downstream validation mechanism.

For the projection-level comparison, we use an attack-timescale-informed reference horizon for each environment, selected from the documented temporal extent of the attack campaign\footref{fn:darpa-e3}: $\Delta t=60$ min for \textsc{Cadets E3} and $\Delta t=40$ min for \textsc{Theia E3}, reflecting the different timescales over which post-alert attack behaviour unfolds in each environment. These horizons are representative configurations for comparing comprehension- and projection-level CSA, not choices optimised for validation performance; Section~\ref{sec:rq2} systematically examines the effect of varying $\Delta t$.

\begin{table}[t]
\centering
\caption{Alert validation across Comprehension (C) and Projection (O) levels on \textsc{Cadets E3} and
\textsc{Theia E3} using $\Delta t=60$ and $40$ min. Higher Recall@$K$ and lower MPR are better.}
\label{tab:rq1_csa_performance}
\renewcommand{\arraystretch}{1.08}
\setlength{\tabcolsep}{3.2pt}
\footnotesize
\begin{tabular}{ll l ccc r}
\toprule
\textbf{Dataset} &
\textbf{CSA level} &
\textbf{Evidence Use} &
\multicolumn{3}{c}{\textbf{Recall@$K$} $\uparrow$} &
\textbf{MPR $\downarrow$} \\
\cmidrule(lr){4-6}
& & & \textbf{5} & \textbf{10} & \textbf{20} & \\
\midrule

\multirow{6}{*}{\textsc{Cadets}}
& \multirow{3}{*}{C}
& Queue      & 0.286 & 0.286 & 0.571 & 17.57 \\
& & Max.~CSG & 0.143 & 0.286 & 0.571 & 18.00 \\
& & PU-MIL   & 0.429 & 0.571 & 0.714 & 13.57 \\
\cmidrule(lr){2-7}
& \multirow{3}{*}{O$_{60}$}
& Queue      & 0.000 & 0.000 & 0.286 & 36.14 \\
& & Max.~CSG & 0.143 & 0.286 & 0.571 & 20.00 \\
& & PU-MIL   & 0.429 & 0.714 & 0.714 & 12.57 \\

\midrule

\multirow{6}{*}{\textsc{Theia}}
& \multirow{3}{*}{C}
& Queue      & 0.000 & 0.000 & 0.000 & 166.75 \\
& & Max.~CSG & 0.000 & 0.250 & 0.250 & 95.00 \\
& & PU-MIL   & 0.750 & 0.750 & 0.750 & 22.00 \\
\cmidrule(lr){2-7}
& \multirow{3}{*}{O$_{40}$}
& Queue      & 0.000 & 0.000 & 0.500 & 163.00 \\
& & Max.~CSG & 0.000 & 0.000 & 0.250 & 176.75 \\
& & PU-MIL   & 0.500 & 0.500 & 1.000 & 8.50 \\

\bottomrule
\end{tabular}
\end{table}

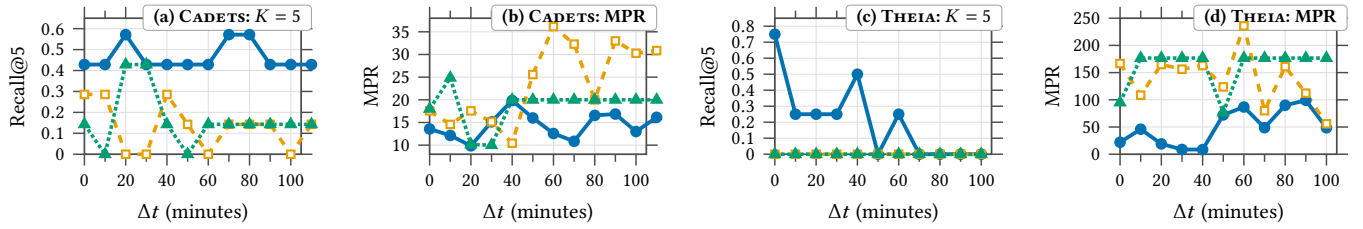
\begin{figure*}[t]
    \centering

    \begin{tikzpicture}

    \begin{groupplot}[
        group style={
            group size=4 by 1,
            horizontal sep=1.7cm,
            group name=rqtwoplots
        },
        width=0.250\textwidth,
        height=0.19\textwidth,
        xmin=0,
        xmax=105,
        xtick={0,20,40,60,80,100},
        minor x tick num=1,
        xlabel={$\Delta t$ (minutes)},
        grid=major,
        major grid style={
            draw=black!13,
            line width=0.35pt
        },
        axis line style={
            draw=black!70,
            line width=0.65pt
        },
        tick style={
            draw=black!70,
            line width=0.55pt
        },
        tick align=outside,
        tick label style={
            font=\footnotesize
        },
        label style={
            font=\small
        },
        xlabel style={
            yshift=0pt
        },
        ylabel style={
            yshift=-1pt
        },
        legend style={
            font=\small,
            draw=none,
            fill=none,
            legend columns=3,
            column sep=0.45cm,
            row sep=0pt,
            /tikz/every even column/.append style={
                column sep=0.06cm
            }
        },
        legend image post style={
            xscale=0.65
        },
        legend cell align={left},
        clip=false
    ]

    \nextgroupplot[
        ylabel={Recall@$5$},
        ymin=0,
        ymax=0.65,
        ytick={0,0.1,0.2,0.3,0.4,0.5,0.6}
    ]

    \addplot[pumil]
    table[
        x=horizon_min,
        y=pumil_r5,
        col sep=comma
    ]{graphs/rq2_projection_with_baselines.csv};

    \addplot[queue]
    table[
        x=horizon_min,
        y=queue_r5,
        col sep=comma
    ]{graphs/rq2_projection_with_baselines.csv};

    \addplot[maxcsg]
    table[
        x=horizon_min,
        y=maxcsg_r5,
        col sep=comma
    ]{graphs/rq2_projection_with_baselines.csv};

    \node[
        draw=black!35,
        fill=white,
        rounded corners=1.5pt,
        inner xsep=4pt,
        inner ysep=1.5pt,
        font=\footnotesize\bfseries,
        anchor=north east
    ]
    at (rel axis cs:1.05,1.10)
    {(a) \textsc{Cadets}: $K=5$};

    \nextgroupplot[
        ylabel={MPR},
        ymin=8,
        ymax=38,
        ytick={10,15,20,25,30,35}
    ]

    \addplot[pumil]
    table[
        x=horizon_min,
        y=pumil_mpr,
        col sep=comma
    ]{graphs/rq2_projection_with_baselines.csv};

    \addplot[queue]
    table[
        x=horizon_min,
        y=queue_mpr,
        col sep=comma
    ]{graphs/rq2_projection_with_baselines.csv};

    \addplot[maxcsg]
    table[
        x=horizon_min,
        y=maxcsg_mpr,
        col sep=comma
    ]{graphs/rq2_projection_with_baselines.csv};

    \node[
        draw=black!35,
        fill=white,
        rounded corners=1.5pt,
        inner xsep=4pt,
        inner ysep=1.5pt,
        font=\footnotesize\bfseries,
        anchor=north east
    ]
    at (rel axis cs:1.05,1.10)
    {(b) \textsc{Cadets}: MPR};

    \nextgroupplot[
        ylabel={Recall@$5$},
        ymin=0,
        ymax=0.85,
        ytick={0,0.1,0.2,0.3,0.4,0.5,0.6,0.7,0.8}
    ]

    \addplot[pumil]
    table[
        x=horizon_min,
        y=pumil_r5,
        col sep=comma
    ]{graphs/rq2_projection_with_baselines_theia.csv};

    \addplot[queue]
    table[
        x=horizon_min,
        y=queue_r5,
        col sep=comma
    ]{graphs/rq2_projection_with_baselines_theia.csv};

    \addplot[maxcsg]
    table[
        x=horizon_min,
        y=maxcsg_r5,
        col sep=comma
    ]{graphs/rq2_projection_with_baselines_theia.csv};

    \node[
        draw=black!35,
        fill=white,
        rounded corners=1.5pt,
        inner xsep=4pt,
        inner ysep=1.5pt,
        font=\footnotesize\bfseries,
        anchor=north east
    ]
    at (rel axis cs:1.05,1.10)
    {(c) \textsc{Theia}: $K=5$};

    \nextgroupplot[
        ylabel={MPR},
        ymin=0,
        ymax=250,
        ytick={0,50,100,150,200,250}
    ]

    \addplot[pumil]
    table[
        x=horizon_min,
        y=pumil_mpr,
        col sep=comma
    ]{graphs/rq2_projection_with_baselines_theia.csv};

    \addplot[queue]
    table[
        x=horizon_min,
        y=queue_mpr,
        col sep=comma
    ]{graphs/rq2_projection_with_baselines_theia.csv};

    \addplot[maxcsg]
    table[
        x=horizon_min,
        y=maxcsg_mpr,
        col sep=comma
    ]{graphs/rq2_projection_with_baselines_theia.csv};

    \node[
        draw=black!35,
        fill=white,
        rounded corners=1.5pt,
        inner xsep=4pt,
        inner ysep=1.5pt,
        font=\footnotesize\bfseries,
        anchor=north east
    ]
    at (rel axis cs:1.05,1.10)
    {(d) \textsc{Theia}: MPR};

    \end{groupplot}
    \end{tikzpicture}
    \caption{
        Effect of projection horizon on alert prioritisation across
        \textsc{Cadets E3} and \textsc{Theia E3}.
        Panels~(a) and~(b) report Recall@$5$ and mean positive rank
        (MPR) for \textsc{Cadets E3}; 
        Panels~(c) and~(d) report the corresponding metrics for
        \textsc{Theia E3}. PU-MIL (\protect\pumilsymbol), Queue (\protect\queuesymbol), and Max-CSG (\protect\maxcsgsymbol) denote the three ranking methods.
        $\Delta t=0$ corresponds to comprehension-level
        CSA, while $\Delta t>0$ incorporates post-alert provenance activity.
        Higher Recall@$5$ and lower MPR indicate better prioritisation.
    }
    \label{fig:rq2-projection-horizon}
\end{figure*}

\paragraph{Comprehension provides alert-discriminative causal context.}
Across both evaluation environments, comprehension-level CSA transforms an isolated alert into behavioural hypotheses describing its causal provenance context. Anomaly-derived reference rankings show that this reconstructed context captures suspicious behaviour; however, anomalousness alone does not indicate whether an alert is attack-related. On \textsc{Cadets E3}, queue- and maximum-CSG anomaly rankings recover 28.6\% and 14.3\% of the attack-confirmed alerts within the top five, respectively. On \textsc{Theia E3}, neither ranking recovers an attack-confirmed alert within the top five, despite maximum-CSG ranking recovering 25\% within the top ten. Thus, reconstructed causal context can expose suspicious attack behaviour without providing a reliable alert-level ordering.

PU-MIL provides a complementary perspective by explicitly learning how this comprehension-level evidence relates to alert validation. Using the same information, it places 42.9\% of the \textsc{Cadets} attack-confirmed alerts and 75\% of the \textsc{Theia} attack-confirmed alerts within the top five, with MPRs of 13.57 and 22.00, respectively. These results should not be interpreted primarily as outperforming the anomaly-derived rankings, since the two mechanisms optimise different objectives. Rather, they show that comprehension-level causal evidence can be explicitly associated with attack-confirmed alerts once a model is trained for alert validation. \textit{In other words, comprehension-level CSA contributes more than a suspiciousness estimate; by reconstructing the causal provenance context associated with an alert, it provides discriminative behavioural evidence for determining whether it is actually attack-related. A learning-based alert-validation model can exploit this to prioritise attack-confirmed alerts more effectively than anomaly severity alone.}

\paragraph{Projection-level CSA improves prioritisation of difficult alerts.} 
Projection-level CSA extends causal reconstruction beyond alert detection time, exposing how the associated behavioural context evolves. Across both environments, the anomaly-derived reference rankings indicate that incorporating projected provenance does not translate directly into stronger anomalousness-based prioritisation. On \textsc{Cadets E3}, extending the observation horizon to 60 minutes reduces queue-anomaly R@20 from 0.571 to 0.286 and increases MPR from 17.57 to 36.14, while maximum-CSG ranking preserves its Recall@$K$ values but increases MPR from 18.00 to 20.00. On \textsc{Theia E3}, the 40-minute horizon leaves queue-level early recall low despite increasing R@20 to 0.500, while maximum-CSG ranking decreases from R@10=0.250 (comprehension) to 0.000 (projection), with MPR rising from 95.00 to 176.75. Projection-level CSA does not inherently make attack-confirmed alerts appear more anomalous. Instead, it contributes subsequent behavioural evidence.

PU-MIL shows that this additional evidence can still inform alert validation. On \textsc{Cadets E3}, projection preserves R@5 and R@20 at 0.429 and 0.714, respectively, while increasing R@10 from 0.571 to 0.714 and reducing MPR from 13.57 to 12.57. Projection therefore selectively refines alert ordering rather than uniformly improving performance across investigation depths. The effect is more pronounced on \textsc{Theia E3}. At the 40-minute horizon, PU-MIL changes from R@5/R@10/R@20 of 0.750/0.750/0.750 at comprehension to 0.500/0.500/1.000, while MPR decreases substantially from 22.00 to 8.50. Although fewer attack-confirmed alerts remain within the stricter top-5 and top-10 cutoffs, all are recovered within the top 20, with a substantial improvement in overall ranking positions. This suggests that projection-level CSA redistributes evidential support across alerts rather than uniformly strengthening already prioritised ones. 
\textit{Projection-level CSA is most valuable when comprehension-level context is insufficient to establish attack relevance, providing future behavioural evidence that helps distinguish difficult attack alerts from benign alternatives.}



\subsection{Impact of Projection Horizon}
\label{sec:rq2}

RQ2 examines how the temporal extent of projection-level CSA influences its usefulness for alert validation. We vary the projection horizon $\Delta t$ from 10 to 100 minutes, with $\Delta t=0$ corresponding to comprehension-level CSA, and evaluate its effect on both \textsc{Cadets E3} and \textsc{Theia E3} using Recall@5 and MPR, with a common Recall@$K$ cutoff across both datasets.
Figure~\ref{fig:rq2-projection-horizon} reports the horizon-dependent behaviour. Panels~(a) and~(b) show Recall@5 and MPR, respectively, for \textsc{Cadets E3}, while Panels~(c) and~(d) report the corresponding metrics for \textsc{Theia E3}. 

\paragraph{Projection utility is inherently horizon-dependent.}
Figure~\ref{fig:rq2-projection-horizon}(a--d) reveals this first through the anomaly-derived reference rankings. Across both datasets, Queue and Max.~CSG rankings vary non-monotonically as $\Delta t$ increases: extending the observation horizon can improve, preserve, or degrade prioritisation of attack-confirmed alerts. Since these rankings directly reflect the provenance reconstructed at each horizon, this behaviour shows that progressively extending the temporal scope does not produce a correspondingly monotonic increase in validation-relevant anomalous evidence. Projection therefore changes the behavioural evidence available for analysis rather than simply accumulating uniformly useful context.

PU-MIL exhibits the same horizon dependence when this information is explicitly learned for alert validation. Its performance does not improve monotonically with $\Delta t$ on either dataset, despite achieving substantial gains at particular intermediate horizons. On \textsc{Cadets E3}, projection can improve both Recall@5 and MPR relative to comprehension-level information, whereas on \textsc{Theia E3} the benefit is more pronounced in the overall ranking: at $\Delta t=40$ min, MPR decreases from 22.00 to 8.50 even though Recall@5 does not exceed its comprehension-level value. Thus, task-specific learning can exploit projection-level information to improve alert validation, but it does not eliminate the underlying temporal sensitivity.
\textit{Both anomaly-based rankings and PU-MIL show that projection-level CSA derives its value not from additional future activity, but from exposing validation-relevant evidence at specific projection horizons.} 

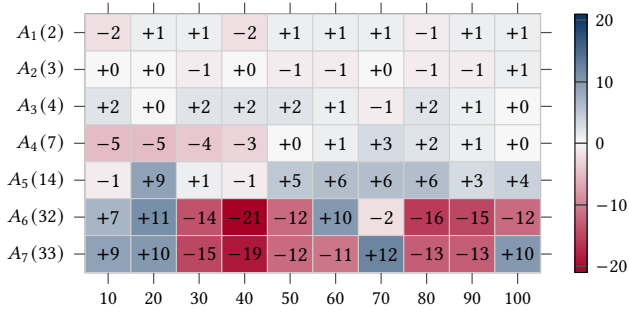
\begin{figure}[t]
    \centering

    \begin{tikzpicture}

    \begin{axis}[
        width=7.6cm,
        height=5.0cm,
        xmin=5,
        xmax=105,
        ymin=0.5,
        ymax=7.5,
        y dir=reverse,
        xtick={10,20,30,40,50,60,70,80,90,100},
        ytick={1,2,3,4,5,6,7},
        yticklabels={
            {$A_1 (2)$},
            {$A_2 (3)$},
            {$A_3 (4)$},
            {$A_4 (7)$},
            {$A_5 (14)$},
            {$A_6 (32)$},
            {$A_7 (33)$}
        },
        tick label style={
            font=\footnotesize
        },
        axis line style={
            draw=black!70,
            line width=0.55pt
        },
        tick style={
            draw=black!70,
            line width=0.45pt
        },
        tick align=outside,
        enlargelimits=false,
        colormap name=rankgain,
        point meta min=-21,
        point meta max=21,
        colorbar,
        colorbar style={
            ytick={-20,-10,0,10,20},
            tick label style={
                font=\footnotesize
            },
            width=0.16cm
        },
        clip=false
    ]

    \addplot[
        matrix plot*,
        mesh/cols=10,
        point meta=explicit,
        draw=black!18,
        line width=0.20pt
    ]
    table[
        x=horizon_min,
        y=alert_row,
        meta=rank_gain,
        col sep=comma
    ]{graphs/rq2_rank_gain_heatmap.csv};

    \addplot[
        only marks,
        mark=none,
        nodes near coords*={
            \pgfmathprintnumber[
                showpos,
                fixed,
                precision=0
            ]{\cellvalue}
        },
        visualization depends on={
            \thisrow{rank_gain}
            \as\cellvalue
        },
        every node near coord/.append style={
            font=\small,
            text=black,
            anchor=center
        }
    ]
    table[
        x=horizon_min,
        y=alert_row,
        col sep=comma
    ]{graphs/rq2_rank_gain_heatmap.csv};

    \end{axis}

    \end{tikzpicture}

    \caption{
        Projection-induced rank shift on
        \textsc{Cadets E3}. Alerts $A_1$--$A_7$ are shown in comprehension-rank order. 
        Positive values improve
        prioritisation; negative values degrade it. 
    }
    \label{fig:rq2-rank-gain}

\end{figure}

\paragraph{Projection affects individual alerts heterogeneously.}
Figure~\ref{fig:rq2-rank-gain} examines the alert rank transitions underlying the aggregate \textsc{Cadets E3} results. The effect of varying $\Delta t$ is not uniform across attack-confirmed alerts. Alerts highly prioritised using comprehension-level CSA remain relatively stable across projection horizons, exhibiting only modest rank adjustments. In contrast, lower-ranked alerts exhibit substantially larger non-monotonic rank changes: the same alert may improve at some horizons and worsen at others. Consequently, horizon sensitivity arises primarily from greater ranking volatility in initially lower-ranked alerts rather than from reordering already highly prioritised alerts. \textit{Projection-level CSA has the greatest impact when attack relevance remains uncertain after comprehension-level analysis, whereas alerts already strongly supported by that evidence exhibit little sensitivity to projection horizon.}



\section{Discussion}

This paper investigated whether progressively richer CSA information improves alert validation and how its utility varies with temporal scope. The results show that richer CSA levels can provide validation-relevant evidence beyond anomalousness alone, while the benefits of projection-level CSA depend strongly on the behavioural context captured within the observation horizon. This section discusses the implications for understanding progressive CSA, supporting SOC alert validation, and guiding future research.

\subsection{Understanding Progressive CSA Behaviour} \label{sec:discussion_behavior}

The results in Sections~\ref{sec:rq1}-\ref{sec:rq2} reveal two complementary properties of progressive CSA. First, enriching an alert with causal and subsequent behavioural context provides additional evidence for validation, but behavioural anomalousness alone does not uniformly reflect this contribution. Instead, PU-MIL demonstrates that the generated CSA information captures validation-relevant signals beyond those reflected in anomalousness scores. Second, the usefulness of projection-level information depends strongly on the horizon. Across both \textsc{Cadets E3} and \textsc{Theia E3}, extending $\Delta t$ does not produce a monotonic improvement in alert prioritisation, because increasing $\Delta t$ does not simply provide more of the same evidence. Instead, it changes which portion of the post-alert system evolution becomes available for reasoning. Different horizons can therefore expose behaviour with different relevance to the alert under validation. The corresponding non-monotonic behaviour of the anomaly-derived reference rankings supports this interpretation: horizon sensitivity is already present in the reconstructed provenance evidence and is subsequently inherited by task-specific learning. The value of projection thus depends on the relevance of the behavioural context exposed within the observation horizon rather than on the amount of additional context alone. 


\subsection{Implications for SOC Alert Validation}
\label{sec:discussion_soc}

The findings suggest that progressive CSA can support evidence-directed allocation of SOC investigation effort, while raising an important operational question: when should an alert be escalated to a richer CSA level? Our results indicate that comprehension-level evidence may suffice for some alerts, whereas others benefit from projection-level CSA. However, the current framework does not make this determination. This motivates adaptive progression policies in which the strength or uncertainty of validation evidence determines whether further CSA enrichment is warranted, rather than requiring every alert to undergo the full analysis pipeline.

A related challenge concerns how far the projection should extend, if at all. The horizon sensitivity observed in Section~\ref{sec:rq2} rules out a simple ``longer is better'' strategy. A fixed observation horizon may include irrelevant future context or fail to capture information important for alert validation, whether inferred through prediction or revealed through subsequent system behaviour. In operational settings, the projection horizon should be an adaptive decision that balances the expected validation value of additional evidence against the cost of acquiring, processing, or reasoning over it. Learning when to invoke projection, how far it should extend, and when sufficient evidence has been acquired remains a key challenge for operationalising progressive CSA in practice.

\subsection{Limitations and Future Directions}
\label{sec:discussion_limitations}

Our findings should be interpreted in light of three limitations. First, end-to-end evaluation is constrained to datasets supported by all components of the progressive CSA pipeline. As a result, experiments are limited to \textsc{Cadets E3} and \textsc{Theia E3}. Although these datasets represent distinct provenance environments and attack campaigns, they are insufficient to establish that the observed behaviour generalises across diverse SOC environments.
%
%
Second, projection is based on subsequently observed provenance activity rather than forecasts of future attacker behaviour. While this enables a controlled evaluation of the value of post-alert behavioural evidence for validation, it does not establish that such evidence can be predicted accurately at alert time. Future work should investigate attacker-behaviour forecasting to determine whether predicted future context yields similar benefits. 
Finally, the framework relies on predefined projection horizons and fixed CSA progression levels. 
The performance variation observed across CSA levels and projection horizons suggests that these choices influence both validation effectiveness and operational efficiency. Future work should investigate adaptive progression and stopping policies that determine when projection is needed and how much future context to acquire based on available evidence and uncertainty.
Collectively, these limitations motivate future research on adaptive and predictive progressive CSA frameworks for alert validation. 
\section{Conclusion}
\label{sec:conclusion}

This work investigated how progressively richer CSA supports alert validation in SOCs. We introduced a unified framework that integrates perception-, comprehension-, and projection-level awareness, operationalised through provenance-based analysis and weakly supervised alert validation. Experimental evaluation demonstrates that comprehension-level CSA consistently improves validation by providing causal attack context beyond the initial alert. In contrast, the benefits of projection-level CSA are conditional rather than cumulative. Because projection-level awareness is inherently predictive, its effectiveness depends on the temporal horizon over which subsequent behaviour is anticipated. These findings challenge the assumption that progressively richer CSA necessarily leads to better validation outcomes. Instead, they reframe alert validation as an awareness-allocation problem, where additional evidence should be acquired and presented only when it actively improves decision-making. Future SOC assistance systems should move beyond providing ever-richer context towards adaptively delivering the specific level and type of awareness that an alert demands. Future work should focus on learning when higher levels of CSA are required, which evidence is most informative for different alert types, and when awareness is sufficient to support timely and accurate analyst decisions. 




\bibliographystyle{ACM-Reference-Format}
\bibliography{references}
\appendix
\section*{Open Science}
To support reproducibility and facilitate future research, we plan to publicly release the implementation of our progressive CSA-aware PU-MIL framework and the trained \textsc{Captain} model for \textsc{Cadets-E5} upon submission to a peer-reviewed conference.

\section{Alternative Realisations of Progressive CSA Generation}
\label{app:alternative_realisation}
\subsection{Perception-Level CSA Generation}
\label{app:perception_realisation}

Perception-level CSA generation can be realised through both knowledge-driven and data-driven provenance analysis paradigms. As shown in Table~\ref{tab:perception_realisation}, the primary distinction lies in the source of knowledge used to infer suspiciousness from provenance data.

Knowledge-driven approaches encode security semantics directly into provenance reasoning, for example, by propagating integrity or trust labels through provenance graphs to identify suspicious interactions~\cite{hossain2017sleuth,hossain2020combating,li2025incorporating}. In contrast, data-driven approaches infer suspiciousness from patterns learned directly from provenance data, such as identifying anomalous interactions using learned provenance representations~\cite{cheng2024kairos,jiang2025orthrus}. Both paradigms ultimately generate perception-level awareness of suspicious entities and interactions, but they expose different trade-offs for downstream CSA construction.

\begin{table*}[t]
\centering
\caption{Alternative realisations of perception-level CSA generation in provenance-based intrusion detection systems.}
\label{tab:perception_realisation}
\small
\begin{tabular}{p{2cm} p{3cm} p{3.2cm} p{3cm} p{4.5cm}}
\toprule
\textbf{Approach}
&
\textbf{Knowledge Source}
&
\textbf{Inference Mechanism}
&
\textbf{Representative Systems}
&
\textbf{Implications for Progressive CSA Construction}
\\
\midrule

Knowledge-Driven
&
Explicit security knowledge encoded into provenance reasoning
&
Reasoning over security semantics associated with provenance activity
&
SLEUTH~\cite{hossain2017sleuth},
MORSE~\cite{hossain2020combating},
CAPTAIN~\cite{li2025incorporating}
&
Preserves causal interpretability and provenance semantics, facilitating downstream CSA expansion, summarisation, and reasoning
\\

Data-Driven
&
Patterns learned directly from provenance activity
&
Identifying deviations from learned provenance behaviour
&
KAIROS~\cite{cheng2024kairos},
ORTHRUS~\cite{jiang2025orthrus}
&
Captures complex attack behaviours and previously unseen activity, but the resulting evidence may be less directly interpretable
\\

\bottomrule
\end{tabular}
\end{table*}
Knowledge-driven approaches provide strong causal interpretability because suspiciousness remains explicitly associated with provenance entities and interactions throughout the detection process. In contrast, data-driven approaches can capture complex dependencies and previously unseen attack behaviours but often rely on learned representations whose security semantics may be less transparent. Consequently, knowledge-driven approaches provide a particularly suitable foundation for progressive CSA construction, where higher CSA levels depend on preserving and reasoning over the provenance context surrounding suspicious activity.

\subsection{Comprehension-Level CSA Generation}
\label{app:comprehension_realisation}

Comprehension-level CSA generation can be realised through several provenance analysis paradigms that differ in how they reconstruct attack-relevant behavioural context from suspicious system activity. As shown in Table~\ref{tab:comprehension_realisation}, the primary distinction lies in the mechanism used to construct causally meaningful provenance structures surrounding an alert. Seed-driven approaches reconstruct attack context by expanding causal dependencies outward from a suspicious event, alert, or detection point, recovering the provenance activity directly connected to the initial suspicious behaviour. Knowledge-guided approaches reconstruct attack context by constraining provenance reconstruction using predefined security knowledge, such as attack templates, threat intelligence, or MITRE ATT\&CK tactics, techniques, and procedures (TTPs), thereby retaining only security-relevant provenance activity. Learning-based approaches reconstruct attack context by learning anomalous behavioural patterns directly from provenance data, identifying and summarising suspicious provenance activity into compact behavioural representations without relying on manually engineered security knowledge. Although these approaches differ in their reconstruction mechanisms, they all seek to generate causally meaningful behavioural representations that extend perception-level alerts with comprehension-level causal awareness.

\begin{table*}[t]
\centering
\caption{Alternative realisations of comprehension-level CSA generation in provenance-based security analysis.}
\label{tab:comprehension_realisation}
\renewcommand{\arraystretch}{1.2}
\small
\begin{tabular}{p{1.8cm}p{4cm}p{2.5cm}p{3.4cm}p{4.0cm}}
\toprule
\textbf{Approach} &
\textbf{Core Principle} &
\textbf{Representative Systems} &
\textbf{Strengths} &
\textbf{Implications for Progressive CSA Construction} \\
\midrule

Seed-Driven Dependency Expansion &
Expand provenance dependencies from suspicious seed events &
BackTracker~\cite{king2005backtracking},
NoDoze~\cite{hassan2019nodoze} &
Preserve causal relationships and provide comprehensive attack context around suspicious activity. &
May reconstruct large provenance graphs that contain substantial benign but causally reachable activity, necessitating additional pruning or ranking mechanisms. \\

\midrule

Knowledge-Guided Provenance Reconstruction &
Apply expert knowledge, attack templates, threat intelligence, TTPs &
POIROT~\cite{milajerdi2019poirot},
HOLMES~\cite{milajerdi2019holmes} &
Produces security-grounded attack context with high semantic interpretability. &
Depends on manually engineered attack knowledge, curated threat intelligence, and expert-defined rules, limiting adaptability to previously unseen attacks. \\

\midrule

Learning-Based Provenance Summarisation &
Learn anomalous provenance behaviour and summarise into compact graphs &
KAIROS~\cite{cheng2024kairos}, ORTHRUS~\cite{jiang2025orthrus} &
Generates compact, scalable behavioural summaries while reducing reliance on handcrafted attack knowledge. &
Effectiveness depends on the quality and representativeness of training data and learned behavioural models. \\

\bottomrule
\end{tabular}
\end{table*}

Seed-driven and knowledge-guided approaches preserve strong causal relationships by explicitly reconstructing provenance dependencies surrounding suspicious activity. Seed-driven reconstruction maximises causal completeness but often introduces large amounts of benign provenance context, whereas knowledge-guided reconstruction improves security relevance by constraining reconstruction using expert knowledge. However, both approaches ultimately depend on predefined expansion criteria or manually engineered security semantics, which may limit scalability and generalisability across diverse attack scenarios.

In contrast, learning-based provenance summarisation reconstructs comprehension-level CSA by identifying anomalous behavioural patterns directly from provenance activity rather than relying on predefined dependency expansion or handcrafted attack knowledge. Consequently, the resulting provenance summaries are typically more compact while still preserving the causal relationships necessary for causal awareness. This provides a particularly suitable foundation for progressive CSA construction, where higher CSA levels require compact behavioural representations that can be efficiently extended, integrated, and subsequently exploited during downstream alert validation.

\subsection{Projection-Level CSA Generation}
\label{appendix:projection_generation}

Projection-level CSA generation extends analysts' awareness beyond the current attack state to anticipate how suspicious behaviour may subsequently evolve. Compared with perception- and comprehension-level CSA generation, relatively few approaches have explicitly addressed projection-level CSA in provenance-based security analysis. Existing work predominantly generates projection-level CSA by forecasting the evolution of future attacks from the currently observed attack context. As summarised in Table~\ref{tab:projection_generation}, these approaches fall broadly into probabilistic forecasting and temporal sequence learning.

\begin{table*}[t]
\centering
\caption{Alternative realisations of projection-level CSA generation.}
\label{tab:projection_generation}
\renewcommand{\arraystretch}{1.2}
\small
\begin{tabular}{p{1.8cm}p{4cm}p{2.5cm}p{3.4cm}p{4.0cm}}
\toprule
\textbf{Approach} &
\textbf{Core Principle} &
\textbf{Representative Work} &
\textbf{Strengths} &
\textbf{Limitations} \\
\midrule

Probabilistic Forecasting &
Predicts future attack evolution using probabilistic models that estimate future attack states from observed attack behaviour. &
Liu et al.~\cite{liu2023causal} &
Provides explicit reasoning about how future attacks progress under uncertainty. &
Forecast accuracy depends on the underlying probabilistic model, and uncertainty increases over longer prediction horizons. \\

\midrule

Temporal Sequence Learning &
Learns temporal attack evolution directly from historical attack or provenance sequences to forecast future attack behaviour. &
EFI~\cite{zhu2024nip} &
Captures the complex temporal evolution of attacks while preserving provenance semantics. &
Requires representative attack data and remains susceptible to long-horizon prediction errors. \\

\midrule

Retrospective Future-Context Construction &
Constructs projection-level CSA by extending provenance analysis beyond the alert-detection point using subsequently observed future system activity rather than explicitly forecasting future attack behaviour. &
This work &
Preserves causal consistency while avoiding uncertainty associated with speculative future-state prediction. &
Provides retrospective rather than real-time projection. \\

\bottomrule
\end{tabular}
\end{table*}


Probabilistic forecasting estimates future attack evolution using probabilistic reasoning over the currently observed attack context, whereas temporal sequence-learning approaches learn future attack behaviour directly from historical provenance sequences. Although both approaches enable the generation of projection-level CSA information, explicit attack forecasting remains inherently challenging because future system behaviour cannot be directly observed at prediction time. Consequently, forecasting performance is affected by uncertainty in future attack evolution, semantic gaps between high-level attack objectives and the low-level provenance activity, scarcity of representative attack data, and cumulative prediction errors over extended forecasting horizons~\cite{zhu2024nip}.


Retrospective future-context construction provides an alternative realisation by extending the provenance observation horizon beyond the alert-detection point to incorporate subsequently observed system activity. Rather than forecasting future attack behaviour, this approach progressively reveals future provenance context while preserving causal continuity, as in comprehension-level provenance analysis. Although retrospective future-context construction does not provide real-time attack prediction, it offers a causally grounded mechanism for constructing projection-level CSA without the uncertainty associated with speculative future-state forecasting. Consequently, it provides a suitable foundation for progressively integrating projection-level CSA into downstream CSA-aware alert validation.

\clearpage
\twocolumn[
\vspace{0pt}
]
\section{Extended Projection-Horizon Results}
\label{app:extended-rq2}

\begin{figure*}[!t]
    \centering

    \begin{tikzpicture}

    \begin{groupplot}[
        group style={
            group size=3 by 1,
            horizontal sep=1.20cm
        },
        width=0.30\textwidth,
        height=0.225\textwidth,
        xmin=0,
        xmax=100,
        ymin=0,
        ymax=1.02,
        xtick={0,20,40,60,80,100},
        ytick={0,0.2,0.4,0.6,0.8,1.0},
        xlabel={$\Delta t$ (min)},
        tick label style={
            font=\scriptsize
        },
        label style={
            font=\scriptsize
        },
        title style={
            font=\small
        },
        axis line style={
            draw=black!70,
            line width=0.50pt
        },
        tick style={
            draw=black!70,
            line width=0.40pt
        },
        tick align=outside,
        grid=major,
        major grid style={
            draw=black!12,
            line width=0.30pt
        },
        enlargelimits=false,
        clip=true
    ]


    \nextgroupplot[
        title={(a) $K=10$},
        ylabel={Recall@$K$}
    ]

    \addplot[pumil]
    table[
        x=horizon_min,
        y=pumil_r10,
        col sep=comma
    ]{graphs/rq2_extended_cadets.csv};

    \addplot[queue]
    table[
        x=horizon_min,
        y=queue_r10,
        col sep=comma
    ]{graphs/rq2_extended_cadets.csv};

    \addplot[maxcsg]
    table[
        x=horizon_min,
        y=maxcsg_r10,
        col sep=comma
    ]{graphs/rq2_extended_cadets.csv};


    \nextgroupplot[
        title={(b) $K=20$}
    ]

    \addplot[pumil]
    table[
        x=horizon_min,
        y=pumil_r20,
        col sep=comma
    ]{graphs/rq2_extended_cadets.csv};

    \addplot[queue]
    table[
        x=horizon_min,
        y=queue_r20,
        col sep=comma
    ]{graphs/rq2_extended_cadets.csv};

    \addplot[maxcsg]
    table[
        x=horizon_min,
        y=maxcsg_r20,
        col sep=comma
    ]{graphs/rq2_extended_cadets.csv};


    \nextgroupplot[
        title={(c) $K=30$},
        legend style={
            at={(0.5,-0.42)},
            anchor=north,
            legend columns=3,
            draw=none,
            font=\scriptsize,
            column sep=0.18cm
        }
    ]

    \addplot[pumil]
    table[
        x=horizon_min,
        y=pumil_r30,
        col sep=comma
    ]{graphs/rq2_extended_cadets.csv};

    \addplot[queue]
    table[
        x=horizon_min,
        y=queue_r30,
        col sep=comma
    ]{graphs/rq2_extended_cadets.csv};

    \addplot[maxcsg]
    table[
        x=horizon_min,
        y=maxcsg_r30,
        col sep=comma
    ]{graphs/rq2_extended_cadets.csv};

    \end{groupplot}

    \end{tikzpicture}

    \vspace{0.20cm}

    \caption{
        Extended Recall@$K$ results for \textsc{Cadets E3} across
        projection horizons at investigation budgets
        $K\in\{10,20,30\}$.
        $\Delta t=0$ corresponds to comprehension-level CSA, while
        $\Delta t>0$ incorporates projection-level CSA.
        PU-MIL (\protect\pumilsymbol), Queue (\protect\queuesymbol), and
        Max.~CSG (\protect\maxcsgsymbol) denote the three ranking methods.
    }
    \label{fig:app-cadets-extended}

\end{figure*}
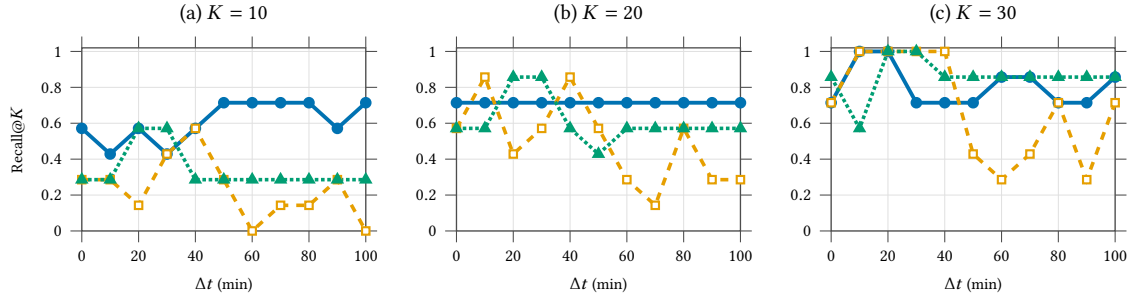

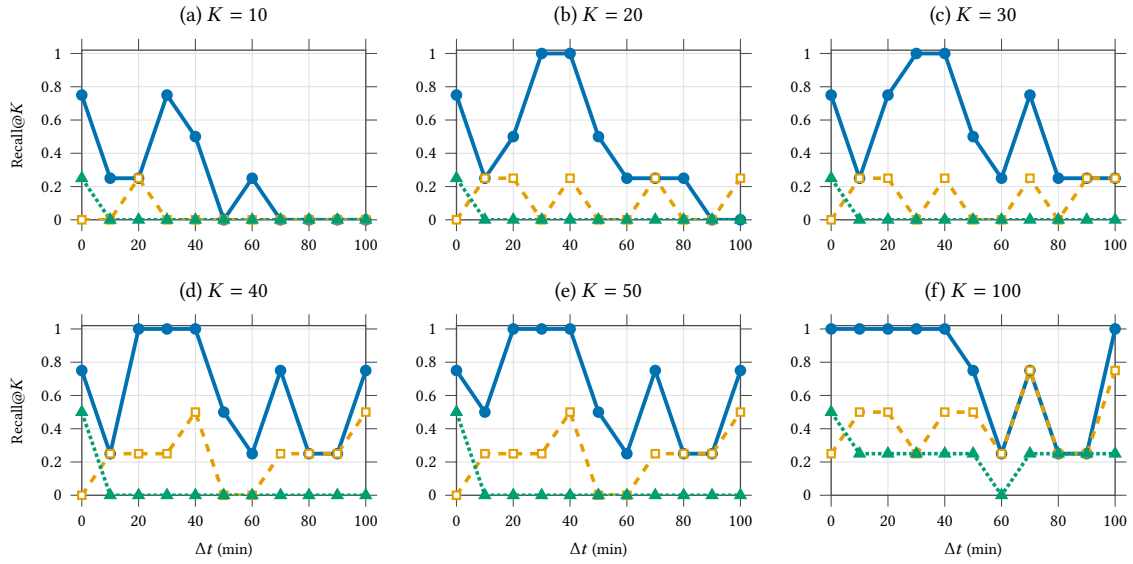
\begin{figure*}[!t]
    \centering

    \begin{tikzpicture}

    \begin{groupplot}[
        group style={
            group size=3 by 2,
            horizontal sep=1.20cm,
            vertical sep=1.40cm
        },
        width=0.30\textwidth,
        height=0.215\textwidth,
        xmin=0,
        xmax=100,
        ymin=0,
        ymax=1.02,
        xtick={0,20,40,60,80,100},
        ytick={0,0.2,0.4,0.6,0.8,1.0},
        tick label style={
            font=\scriptsize
        },
        label style={
            font=\scriptsize
        },
        title style={
            font=\small
        },
        axis line style={
            draw=black!70,
            line width=0.50pt
        },
        tick style={
            draw=black!70,
            line width=0.40pt
        },
        tick align=outside,
        grid=major,
        major grid style={
            draw=black!12,
            line width=0.30pt
        },
        enlargelimits=false,
        clip=true
    ]



    \nextgroupplot[
        title={(a) $K=10$},
        ylabel={Recall@$K$}
    ]

    \addplot[pumil]
    table[
        x=horizon_min,
        y=pumil_r10,
        col sep=comma
    ]{graphs/rq2_extended_theia.csv};

    \addplot[queue]
    table[
        x=horizon_min,
        y=queue_r10,
        col sep=comma
    ]{graphs/rq2_extended_theia.csv};

    \addplot[maxcsg]
    table[
        x=horizon_min,
        y=maxcsg_r10,
        col sep=comma
    ]{graphs/rq2_extended_theia.csv};


    \nextgroupplot[
        title={(b) $K=20$}
    ]

    \addplot[pumil]
    table[
        x=horizon_min,
        y=pumil_r20,
        col sep=comma
    ]{graphs/rq2_extended_theia.csv};

    \addplot[queue]
    table[
        x=horizon_min,
        y=queue_r20,
        col sep=comma
    ]{graphs/rq2_extended_theia.csv};

    \addplot[maxcsg]
    table[
        x=horizon_min,
        y=maxcsg_r20,
        col sep=comma
    ]{graphs/rq2_extended_theia.csv};


    \nextgroupplot[
        title={(c) $K=30$}
    ]

    \addplot[pumil]
    table[
        x=horizon_min,
        y=pumil_r30,
        col sep=comma
    ]{graphs/rq2_extended_theia.csv};

    \addplot[queue]
    table[
        x=horizon_min,
        y=queue_r30,
        col sep=comma
    ]{graphs/rq2_extended_theia.csv};

    \addplot[maxcsg]
    table[
        x=horizon_min,
        y=maxcsg_r30,
        col sep=comma
    ]{graphs/rq2_extended_theia.csv};



    \nextgroupplot[
        title={(d) $K=40$},
        ylabel={Recall@$K$},
        xlabel={$\Delta t$ (min)}
    ]

    \addplot[pumil]
    table[
        x=horizon_min,
        y=pumil_r40,
        col sep=comma
    ]{graphs/rq2_extended_theia.csv};

    \addplot[queue]
    table[
        x=horizon_min,
        y=queue_r40,
        col sep=comma
    ]{graphs/rq2_extended_theia.csv};

    \addplot[maxcsg]
    table[
        x=horizon_min,
        y=maxcsg_r40,
        col sep=comma
    ]{graphs/rq2_extended_theia.csv};


    \nextgroupplot[
        title={(e) $K=50$},
        xlabel={$\Delta t$ (min)}
    ]

    \addplot[pumil]
    table[
        x=horizon_min,
        y=pumil_r50,
        col sep=comma
    ]{graphs/rq2_extended_theia.csv};

    \addplot[queue]
    table[
        x=horizon_min,
        y=queue_r50,
        col sep=comma
    ]{graphs/rq2_extended_theia.csv};

    \addplot[maxcsg]
    table[
        x=horizon_min,
        y=maxcsg_r50,
        col sep=comma
    ]{graphs/rq2_extended_theia.csv};


    \nextgroupplot[
        title={(f) $K=100$},
        xlabel={$\Delta t$ (min)},
        legend style={
            at={(0.5,-0.42)},
            anchor=north,
            legend columns=3,
            draw=none,
            font=\scriptsize,
            column sep=0.18cm
        }
    ]

    \addplot[pumil]
    table[
        x=horizon_min,
        y=pumil_r100,
        col sep=comma
    ]{graphs/rq2_extended_theia.csv};

    \addplot[queue]
    table[
        x=horizon_min,
        y=queue_r100,
        col sep=comma
    ]{graphs/rq2_extended_theia.csv};

    \addplot[maxcsg]
    table[
        x=horizon_min,
        y=maxcsg_r100,
        col sep=comma
    ]{graphs/rq2_extended_theia.csv};

    \end{groupplot}

    \end{tikzpicture}

    \vspace{0.20cm}

    \caption{
        Extended Recall@$K$ results for \textsc{Theia E3} across
        projection horizons at investigation budgets
        $K\in\{10,20,30,40,50,100\}$.
        PU-MIL (\protect\pumilsymbol), Queue (\protect\queuesymbol), and
        Max.~CSG (\protect\maxcsgsymbol) denote the three ranking methods.
        Higher Recall@$K$ indicates better recovery of attack-confirmed
        alerts within the corresponding investigation budget.
    }
    \label{fig:app-theia-extended}

\end{figure*}

In Section~\ref{sec:rq1} evaluation focuses on selected operating points to characterise how the usefulness of projection-level CSA varies with the projection horizon. For completeness, this appendix extends the analysis to broader analyst investigation budgets. Figure~\ref{fig:app-cadets-extended} reports Recall@$K$ for \textsc{Cadets E3} at $K\in\{10,20,30\}$, while Figure~\ref{fig:app-theia-extended} provides a broader analysis for \textsc{Theia E3} at $K\in\{10,20,30,40,50,100\}$.

The extended results are consistent with the principal observation from the main RQ2 analysis rather than indicating a different behaviour at larger investigation budgets. In particular, increasing the projection horizon does not produce a uniformly monotonic
improvement in Recall@$K$. Improvements, degradations, and relatively stable regions remain dependent on both the projection horizon and the investigation budget. Thus, the horizon sensitivity observed at the operating points reported in the main evaluation is not confined to the highest-priority portion of the ranking; it remains visible when progressively larger portions of the prioritised alert queue are considered.

At the same time, increasing $K$ progressively absorbs rank redistributions that are consequential under stricter investigation budgets. As a result, differences between horizons and ranking methods can diminish at larger $K$ as more attack-confirmed alerts fall within the investigated set. This complements the main RQ2 finding: projection-level CSA does not simply become more useful as additional future activity is observed. Rather, its contribution depends on which validation-relevant behavioural evidence is exposed at a given horizon and where the affected alerts subsequently fall within the
prioritised investigation queue.


\end{document}